\documentclass[conference]{IEEEtran}

\usepackage{graphicx}
\usepackage{subfigure}
\usepackage{color}
\usepackage{xspace,makecell,array,booktabs}
\usepackage{tipa}
\usepackage{url}
\usepackage[fleqn]{amsmath}
\usepackage{breqn}
\usepackage{physics}
\usepackage{epstopdf}
\usepackage{citesort}

\let\ipa\textipa
\newcommand{\specificthanks}[1]{\@fnsymbol{#1}}

\newcommand{\name}{{\tt VAuth}\xspace}
\newcommand{\acc}{{accelerometer}\xspace}
\newcommand{\mic}{{microphone}\xspace}

\usepackage{pifont}
\newcommand{\tick}{\ding{52}} 

\newcolumntype{N}{>{\centering\arraybackslash}m{2.8cm}}
\newcolumntype{C}[1]{>{\centering\let\newline\\\arraybackslash\hspace{0pt}}m{#1}}
\newcolumntype{L}[1]{>{\raggedright\let\newline\\\arraybackslash\hspace{0pt}}m{#1}}

\begin{document}


\title{Continuous Authentication for Voice Assistants}
\author{Huan Feng, Kassem Fawaz, and Kang G. Shin\\
	CSE/EECS, University of Michigan\\
	\{huanfeng,kmfawaz,kgshin\}@umich.edu}
\maketitle
\thispagestyle{plain}
\pagestyle{plain}

\begin{abstract}

Voice has become an increasingly popular User Interaction (UI) channel, mainly 
contributing to the ongoing trend of wearables, smart vehicles, and home 
automation systems. Voice assistants such as Siri, Google Now and Cortana, have become 
our everyday fixtures, especially in scenarios where touch interfaces are inconvenient 
or even dangerous to use, such as driving or exercising. Nevertheless, 
the open nature of the voice channel makes voice assistants difficult to secure 
and exposed to various attacks as demonstrated by security researchers. 
In this paper, we present \name, the first system that provides continuous and 
usable authentication for voice assistants. 
We design \name to fit in various widely-adopted wearable devices, such as eyeglasses,
earphones/buds and necklaces, where it collects the body-surface vibrations of the user and 
matches it with the speech signal received by the voice assistant's \mic. 
\name guarantees that the voice assistant executes \textit{only} the commands
that originate from the voice of the owner. 
We have evaluated \name with 18 users and 30 voice commands
and find it to achieve an almost perfect matching 
accuracy with less than 0.1\% false positive rate, regardless of \name's position on the body
and the user's language, accent or mobility. 
\name successfully thwarts different practical attacks, 
such as replayed attacks, mangled voice attacks, or impersonation attacks. 
It also has low energy and latency overheads and is compatible with most existing voice assistants. 

\end{abstract}


\section{Introduction}
\label{sec:introduction}

Siri, Cortana, Google Now, and Alexa are becoming our everyday fixtures. 
Through voice interactions, these and other voice assistants allow us to place phone 
calls, send messages, check emails, schedule appointments, navigate to destinations, 
control smart appliances, and perform banking services.
In numerous scenarios such as cooking, exercising or driving, voice interaction 
is preferable to traditional touch interfaces that are inconvenient or even dangerous 
to use. Furthermore, a voice interface is even essential for the increasingly prevalent 
Internet of Things (IoT) devices that lack touch capabilities~\cite{mit:2014}.

With sound being an open channel, voice as an input mechanism is inherently insecure 
as it is prone to replay, sensitive to noise, and easy to impersonate. 
Existing voice authentication mechanisms, such as Google's ``Trusted Voice" 
and Nuance's ``FreeSpeech" used by banks,\footnote{\small{\url{https://wealth.barclays.com/en_gb/home/international-banking/insight-research/manage-your-money/banking-on-the-power-of-speech.html}}} 
fail to provide the security features for voice assistant systems. 
An adversary can bypass these voice-as-biometric authentication mechanisms by impersonating 
the user's voice or simply launching a replay attack. 
Recent studies have demonstrated that it is possible to inject voice commands remotely 
with mangled voice~\cite{vaidya2015cocaine,carlini2016hidden}, wireless signals~\cite{kasmi2015iemi}, 
or through public radio stations~\cite{npr:2016} without raising the user's attention. 
Even Google warns against its voice authentication feature as being 
insecure,\footnote{\small{When a user tries to enable Trusted Voice on Nexus devices, Google explicitly 
warns that it is less secure than password and can be exploited by the attacker with a very similar voice.}} 
and some security companies~\cite{avg:2014} recommend relinquishing voice interfaces all together 
until security issues are resolved. The implications of attacking voice-assistant 
systems can be severe, ranging from information theft and financial loss~\cite{banking:biometric} 
all the way to inflicting physical harm via unauthorized access to smart appliances and vehicles.

In this paper, we propose \name, a novel system that provides \textit{usable} and 
\textit{continuous} authentication for voice assistant systems. As a wearable 
security token, it supports on-going authentication by matching the user's voice 
with an additional channel that provides physical assurance. \name 
collects the body-surface vibrations of a user via an \acc and continuously 
matches them to the voice commands received by the voice assistant. This way, 
\name guarantees that the voice assistant executes \textit{only} the commands
that originate from the voice of the owner. \name offers the 
following salient features.

\paragraph*{\textbf{Continuous Authentication}} 
\name specifically addresses the problem of continuous authentication of a speaker to a voice-enabled device.
Most authentication mechanisms, including all smartphone-specific ones such as passwords, PINs, 
patterns, and fingerprints, provide security by proving the user's identity before establishing a session. 
They hinge on one underlying assumption: the user retains exclusive control of the device right after the authentication. 
While such an assumption is natural for touch interfaces, it is unrealistic for the case of voice assistants. 
Voice allows access for any third party during a communication session, rendering pre-session 
authentication insufficient.
\name provides ongoing speaker authentication during an entire session by ensuring that every 
speech sample recorded by the voice assistant originates from the speaker's throat.
Thus, \name complements existing mechanisms of initial session authentication and speaker recognition.

\paragraph*{\textbf{Improved Security Features}}

Existing biometric-based authentication approaches tries to reduce 
time-domain signals to a set of vocal features. Regardless of how 
descriptive the features are of the speech signal, they still represent 
a projection of the signal to a reduced-dimension space. Therefore, 
collisions are bound to happen; two different signals can result in 
the same feature vector. For example, Tavish \textit{et al.}~\cite{vaidya2015cocaine}
fabricated mangled voice segments, incomprehensible to a human, but map 
to the same feature vector as a voice command so that they are recognizable 
by a voice assistant. Such attacks weaken the security guarantees provided 
by almost all voice-biometric approaches~\cite{PanjwaniP14}.

In contrast, \name utilizes an instantaneous matching algorithm to compare 
the entire signal from \acc with that of \mic in the time domain.
\name splits both \acc and \mic signals into speech segments and proceeds
to match both signals one segment at a time. It filters
the non-matching segments from the \mic 
signal and only passes the matching ones to the voice assistant.
Our theoretical analysis of \name's matching algorithm (section~\ref{sec:model})
demonstrates that it prevents an attacker from injecting any command even 
when the user is speaking. Moreover, \name overcomes the security problems of 
leaked or stolen voice biometric information, such as voiceprints. 
A voice biometric is a lifetime property of an 
individual, and leaking it renders voice authentication insecure. On the other 
hand, when losing \name for any reason, the user has to just unpair the token 
and pair a new one. 

\paragraph*{\textbf{Usability}}

A user can use \name out-of-the-box as it does not require any user-specific 
training, a drastic departure from existing voice biometric mechanisms. It 
only depends on the instantaneous consistency between the \acc and \mic signals; 
therefore, it is immune to voice changes over time and in different situations, 
such as sickness or tiredness. \name provides its security features as long as 
it touches the user's skin at any position on the facial, throat, and sternum\footnote{The 
sternum is the bone that connects the rib cage; it vibrates as a result of the speech.} 
areas. This allows us to incorporate \name into wearables that people are already using 
on a daily basis, such as eyeglasses, Bluetooth earbuds and necklaces/lockets. Our 
usability survey of 952 individuals revealed that users are willing to accept the 
different configurations of \name, especially when they are concerned about the 
security threats and when \name comes in the forms of which they are already comfortable.

We have implemented a prototype of \name using a commodity \acc and an off-the-shelf Bluetooth 
transmitter. Our implementation is built into the Google Now system in Android, and could 
easily extend to other platforms such as Cortana, Siri, or even phone banking services. To 
demonstrate the effectiveness of \name, we recruited 18 participants and asked each of them 
to issue 30 different voice commands using \name. We repeated the experiments for three 
wearable scenarios: eyeglasses, earbuds and necklace. We found that \name:

\begin{itemize}

\item delivers almost perfect results with more than 97\% detection accuracy and close to 
0 false positives. This indicates most of the commands are correctly authenticated from the 
first trial and \name only matches the command that originates from the owner;

\item works out-of-the-box regardless of variation in accents, mobility patterns (still vs. jogging), 
or even across languages (Arabic, Chinese, English, Korean, Persian);

\item effectively thwarts mangling voice attacks and successfully blocks unauthenticated 
voice commands replayed by an attacker or impersonated by other users; and 

\item incurs negligible latency (an average of 300ms) and energy overhead (requiring 
re-charging only once a week).

\end{itemize}

The rest of the paper is organized as follows. Section~\ref{sec:related} discusses the 
related work while Section~\ref{sec:background} provides the necessary background of 
human speech models. Section~\ref{sec:system-threat-models} states the system and 
threat models and Section~\ref{sec:vauth} details the design and implementation of \name. 
We discuss our matching algorithm in Section~\ref{sec:matching}, and conduct 
phonetic-level analysis on the matching algorithm in Section~\ref{sec:phonetics}. We further 
study the security properties of the matching algorithm in Section~\ref{sec:model} using 
a theoretical model. Section~\ref{sec:evaluation} evaluates \name's effectiveness.  
Section~\ref{sec:discussion} discusses \name's features. Finally, the 
paper concludes with Section~\ref{sec:conclusion}.


\section{Related Work}
\label{sec:related}

\paragraph*{\textbf{Smartphone Voice Assistants}} 
Many researchers have studied the security issues of smartphone voice assistants 
\cite{Diao:2014:YVA:2666620.2666623,kasmi2015iemi,Petracca:2015,vaidya2015cocaine}. 
They have also demonstrated the possibility of injecting commands into voice assistants with 
electromagnetic signals~\cite{kasmi2015iemi} or with a mangled voice that is incomprehensible to 
humans~\cite{vaidya2015cocaine}. These practical attack scenarios motivate us to build 
an authentication scheme for voice assistants. 
Petracca {\em et al.}~\cite{Petracca:2015} proposed a generic protection scheme 
for audio channels by tracking suspicious information flows. 
This solution prompts the user and requires manual review for {\em each} potential voice command.
It thus suffers from the habituation and satisficing drawbacks 
since it interrupts the users from their primary tasks~\cite{felt:2012}.

\paragraph*{\textbf{Voice Authentication}} 
Most voice authentication schemes involve training on the user's voice samples and building 
a voice biometric \cite{baloul2012challenge,cornelius2014vocal,das2008multilingual,kunz2011continuous}. 
The biometric may depend on the user's vocal features or cultural backgrounds and requires rigorous  
training to perform well. There is no theoretical guarantee that they provide good security in general. 
Approaches in this category project the signal to a reduced-dimension space and 
collisions are thus inherent. In fact, most companies adopt these mechanisms for the usability 
benefits and claim they are not as secure as passwords or patterns~\cite{sans:voice:bioemetric}. 
Moreover, for the particular case of voice assistants, they all are subject to simple replay attacks.

\paragraph*{\textbf{Mobile Sensing}} 
Many researchers have studied the potential applications of accelerometers for human behavior 
analysis~\cite{aviv2012practicality,marquardt2011sp,owusu2012accessory,xu2012taplogger}. 
Studies show that it is possible to infer keyboard strokes~\cite{aviv2012practicality}, smartphone touch 
inputs~\cite{xu2012taplogger} or passwords~\cite{aviv2012practicality,owusu2012accessory} 
from acceleration information. 
There are also applications utilizing the correlation between sound and 
vibrations~\cite{lien2015voice,mehta2012mobile} 
for health monitoring purposes. Doctors can thus detect voice disorder without actually collecting 
the user's daily conversations. These studies are very different from ours which focuses on 
{\em continuous} voice assistant security.


\section{background}
\label{sec:background}

We introduce some basic concepts and terminology 
regarding the generation and processing of human speech, which will be 
referenced consistently throughout the paper.

\begin{figure}[t]
     \centering
     \subfigure[Source]{\includegraphics[width=0.43\columnwidth]{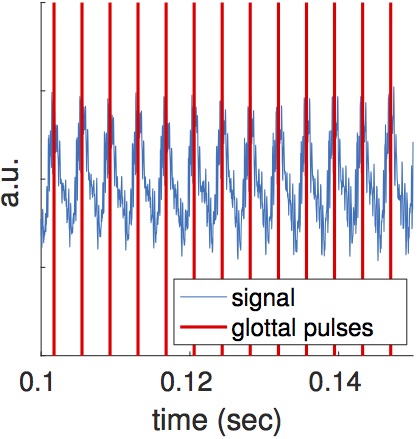}\label{fig:pitch_i}}
     \hfill  
     \subfigure[Filter]{\includegraphics[width=0.43\columnwidth]{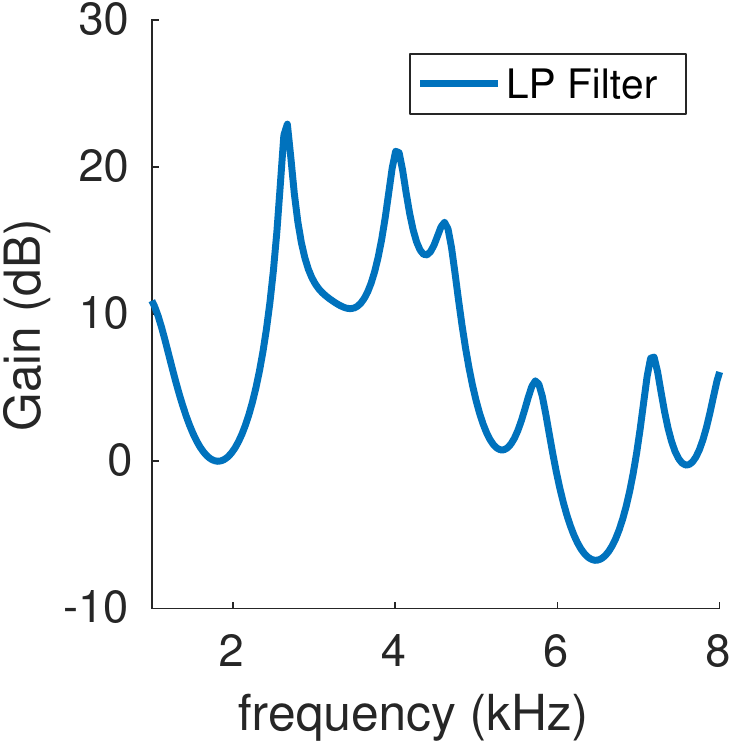}\label{fig:filter_i}}
     \hfill
     \caption{The source--filter model of human speech production using the vowel 
     \texttt{\{i:\}} as an example.}
     \label{fig:speech-model}
\end{figure}

\subsection{Human Speech Model}

The production of human speech is commonly modeled as the combined effect 
of two separate processes~\cite{speech_model}: a voice source (vibration of vocal folds) that generates 
the original signal and a filter (determined by the resonant properties of vocal 
tract including the influence of tongue and lips) that further modulates the signal. 
The output is a shaped spectrum with certain energy peaks, which together maps to 
a specific phoneme (see Fig.~\ref{fig:filter_i} for the vowel \texttt{\{i:\}}
-- the vowel in the word ``see"). 
This process is widely used and referred to as the \textit{source-filter} model.

Fig.~\ref{fig:pitch_i} shows an example of a female speaker pronouncing the 
vowel \texttt{\{i:\}}. The time separating each pair of peaks is the length of 
each glottal pulse (cycle). It also refers to the \textit{instantaneous fundamental frequency} 
($f_0$) variation while the user is speaking, which is the pitch of speaker's voice. 
The value of $f_0$ varies between 80 to 333Hz for a human speaker.
The glottal cycle length (being the inverse
of the fundamental frequency) varies 0.003sec and 0.0125sec.
As the human speaker pronounces 
different phonemes in a particular word, the pitch changes accordingly, 
which becomes an important feature of speaker recognition. We utilize the 
fundamental frequency ($f_0$) as a reference to filter signals that fall outside 
of the human speech range.

\subsection{Speech Recognition and MFCC}

The {\em de facto} standard and probably the most widely used feature for 
speech recognition is Mel-frequency cepstral coefficients 
(MFCC)~\cite{Muda:2010}, which models the way humans perceive sounds. 
In particular, these features are computed on short-term windows 
when the signal is assumed to be stationary. To compute the MFCCs, 
the speech recognition system computes the short-term Fourier 
transform of the signal, then scales the frequency axis to the 
non-linear Mel scale (a set of Mel bands). Then, the Discrete 
Cosine Transform (DCT) is computed on the log of the power spectrum of 
each Mel band. This technique works well in speech recognition 
because it tracks the invariant feature of human speech across 
different users. However, it also opens the door to potential attacks: by 
generating mangled voice segments with the same MFCC feature, an attacker 
can trick the voice assistant into executing specific voice commands 
without drawing any attention from the user.


\section{System and Threat Models}
\label{sec:system-threat-models}

\subsection{System Model}

\name consists of two components. The first is an \acc mounted 
on a wearable device which can be placed on the user's chest, around 
the neck or on the facial area. The second component is an extended voice assistant that issues voice commands 
after correlating and verifying both the \acc signal from the wearable 
device and the \mic signal collected by the assistant. This system is not 
only compatible with smartphone voice assistants such as Siri and Google 
Now, but also applies to voice systems in other domains such as Amazon 
Alexa and phone-based authentication system used by banks. We assume the 
communications between the two components are encrypted.
Attacks to this communication
channel are orthogonal to this work. 
We also assume the wearable device serves as a secure token that the user will not 
share with others. The latter assumption is known as {\em security by 
possession}, which is widely adopted in the security field in the form of 
authentication rings~\cite{Vu:2012}, wristbands~\cite{mare2014zebra},
or RSA SecurID. Thus, the problem of authenticating the wearable token to the user 
is orthogonal to \name and has been addressed elsewhere~\cite{Cornelius:2014}. 
Instead, we focus on the problem of authenticating voice commands, assuming
the existence of a trusted wearable device. 

\subsection{Threat Model}

We consider an attacker who is interested in stealing private information or conducting 
unauthorized operations by exploiting the voice assistant of the target user.
Typically, the attacker tries to hijack the voice assistant of the target 
user and deceive it into executing mal-intended voice commands, such as
sending text messages to premium phone numbers or conducting bank transactions. 
The adversary mounts the attack by interfering with the audio channel. 
This does not assume the attacker has to be physically at 
the same location as the target. It can utilize equipment that can 
generate a sound on its behalf, such as radio channels or high-gain speakers.
Specifically, we consider the following three categories of attack scenarios.

\paragraph*{\textbf{Scenario A -- Stealthy Attack}} 
The attacker attempts to inject either inaudible or incomprehensible voice commands through 
wireless signals~\cite{kasmi2015iemi} or mangled voice commands~\cite{vaidya2015cocaine,carlini2016hidden}. 
This attack is stealthy in the sense that the victim may not even be aware of the on-going 
threat. It is also preferable to the attacker when the victim has physical control or within 
close proximity of the voice assistant.

\paragraph*{\textbf{Scenario B -- Biometric-override Attack}} 
The attacker attempts to inject voice commands~\cite{PanjwaniP14} 
by replaying a previously recorded clip of the victim's voice, or by impersonating the 
user's voice. This attack can have a very low technical barrier: we found that by 
simply mimicking the victim's voice, an attacker can bypass the Trusted Voice feature of 
Google Now within five trials, even when the attacker and the victim are of different genders.

\paragraph*{\textbf{Scenario C -- Acoustic Injection Attack}} 
The attacker can be more advanced, trying to generate a voice that has a direct effect 
on the \acc~\cite{acc:acousticattack}. The intention is to override \name's verification channel with high energy 
vibrations. For example, the attacker can play very loud music which contains embedded 
patterns of voice commands.

\section{VAuth}
\label{sec:vauth}

We now present the high-level design of \name, describe 
our prototype implementation with Google Now, and elaborate on its 
usability aspects.

\subsection{High-Level Overview}

\begin{figure}
	\centering
	\includegraphics[width=0.85\columnwidth]{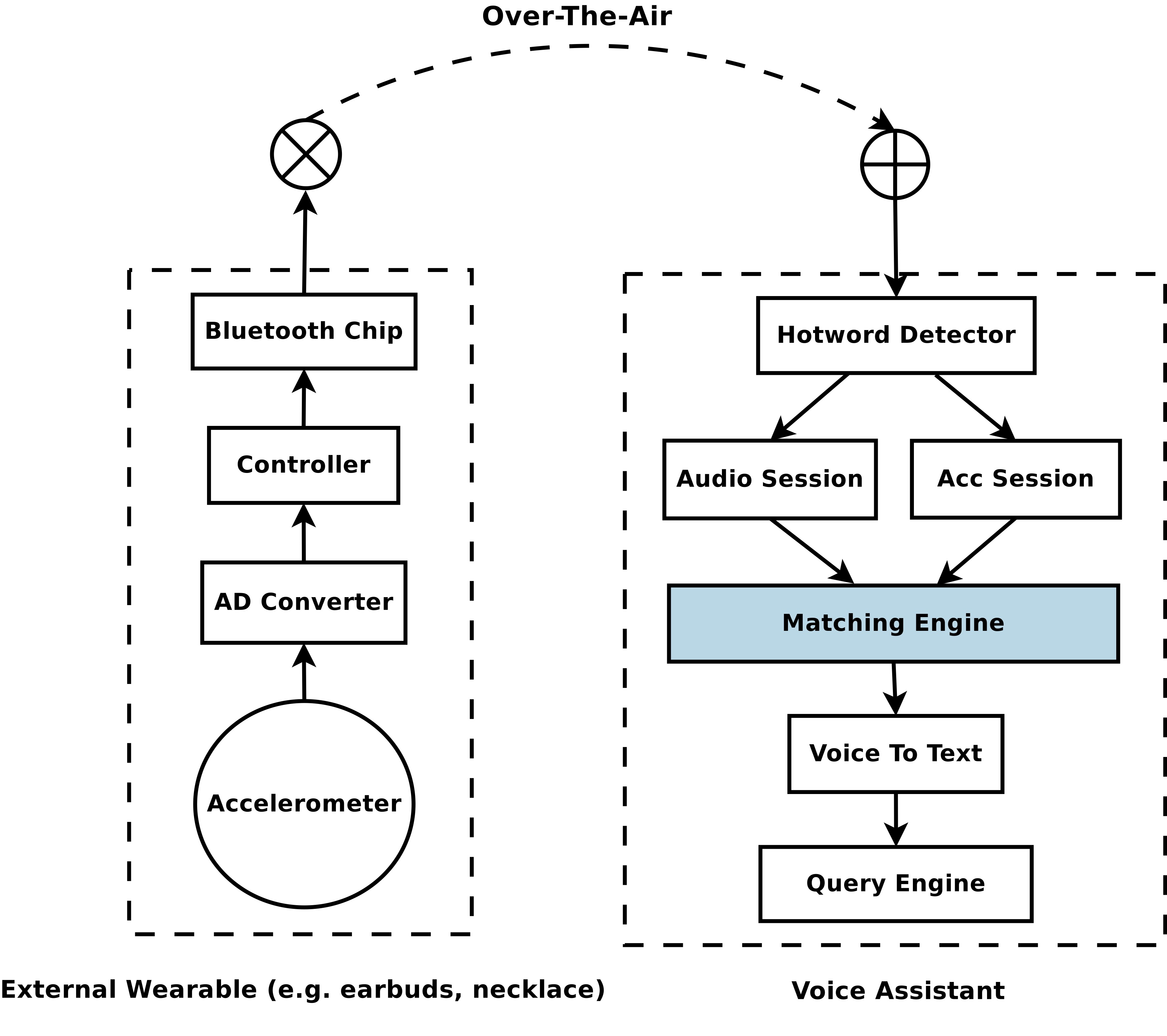}
	\caption{The high-level design of \name, consisting of the wearable and the voice assistant extension.}
	\label{fig:arch}
\end{figure}

\name has two components: (1) a wearable component, responsible for collecting 
and uploading the \acc data, and (2) a voice assistant extension, responsible 
for authenticating and launching the voice commands. 
The first component easily incorporates into existing wearable 
products, such as earbuds/earphones/headsets, eyeglasses, or necklaces/lockets. 
The usability aspect of \name will 
be discussed later in this section
token that the user does not share with others. When a user triggers the voice assistant, 
for example by saying ``OK, Google" or ``Hey, Siri", our voice assistant extension will 
fetch \acc data from the wearable component, correlate it with signals collected 
from \mic and issue the command only when there is a match. Fig.~\ref{fig:arch} depicts the 
information flows in our system. 
To reduce the processing burden on the user's device,
the matching does not take place on the device 
(that runs the voice assistant), but rather at the server side. 
 The communication between the wearable component 
and the voice assistant takes place over Bluetooth BR/EDR~\cite{bt-standard}. 
Bluetooth Classic is an attractive choice as a communication channel, since it has a relatively high 
data rate (up to 2Mbps), is energy-efficient, and enables secure communication through its pairing procedure. 

The design of \name is modular and compatible with most voice assistant systems. 
One can thus customize any component in Fig.~\ref{fig:arch} to optimize functionality, performance 
or usability. Here, we elaborate how to integrate this into an existing voice assistant, 
using Google Now as an example.

\subsection{Prototype}
\label{implementation}

\begin{figure}
	\centering
	\centering
    \subfigure[Wireless]{\includegraphics[width=0.34\columnwidth]{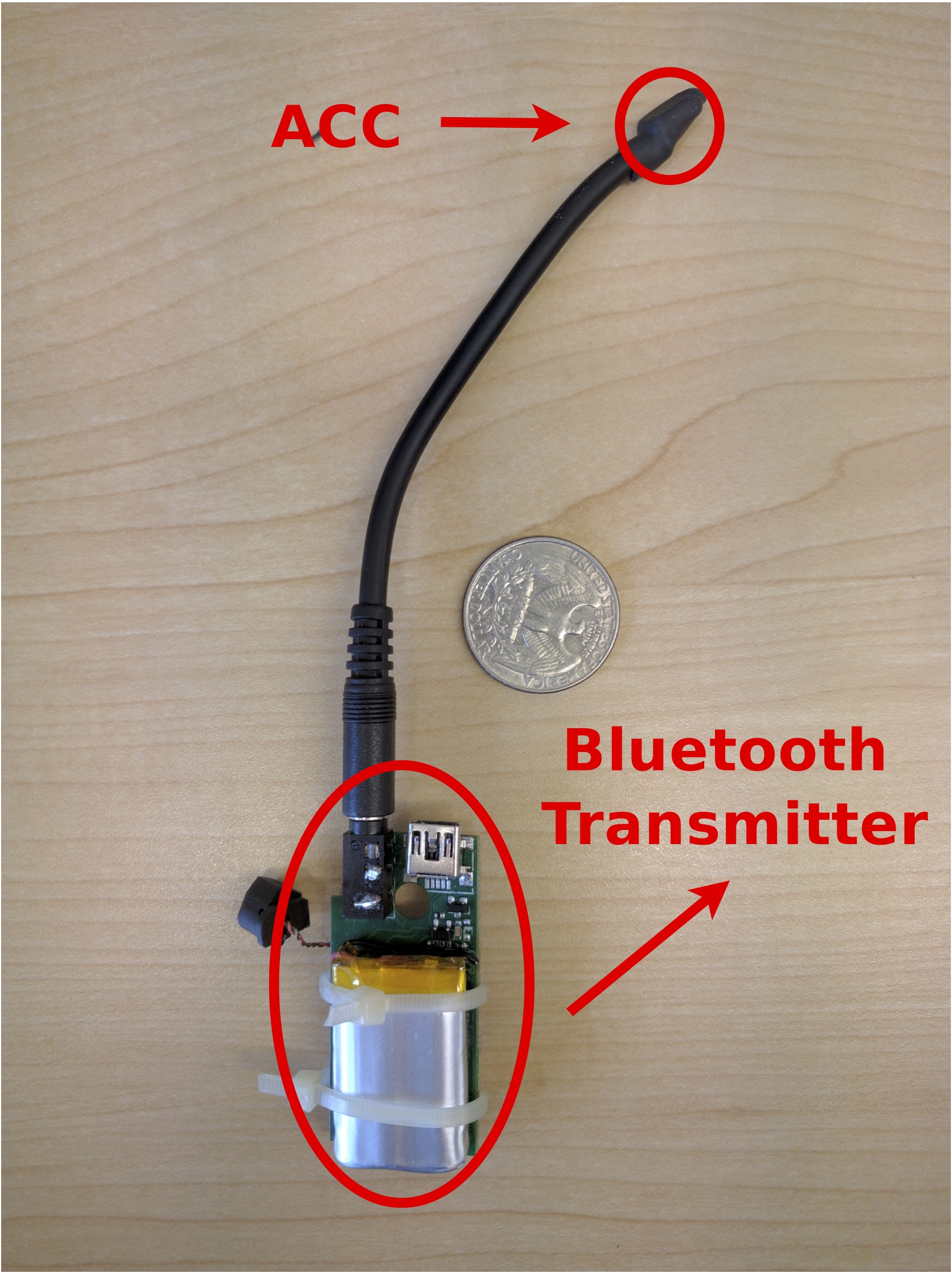}}
    \hfill  
    \subfigure[Eyeglasses]{\includegraphics[width=0.61\columnwidth]{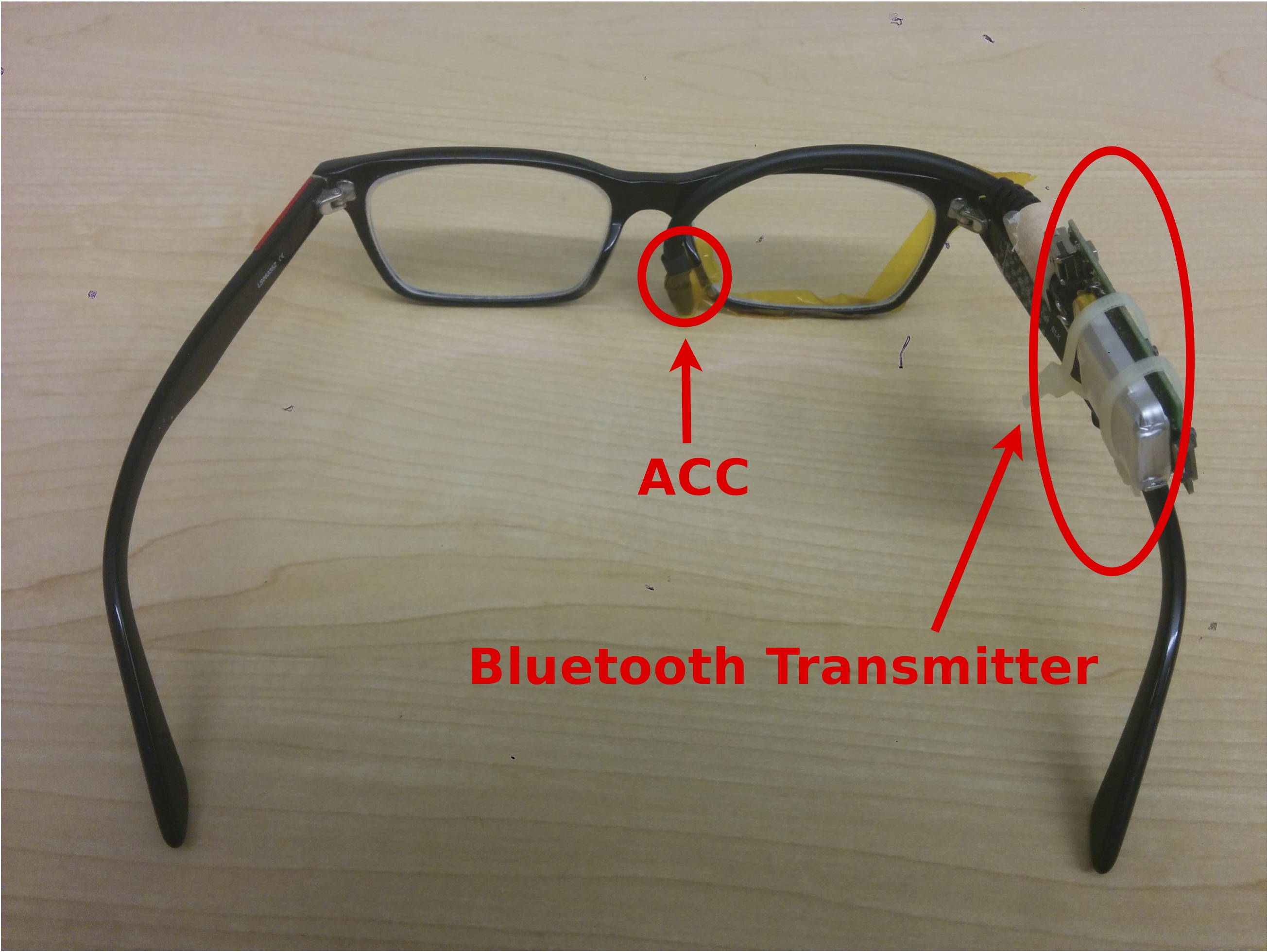}}
    \hfill
	\caption{Our prototype of \name, featuring the \acc chip and Bluetooth transmitter, (a) compared
	to US quarter coin and (b) attached to a pair of eyeglasses belonging to one of the authors.}
	\label{fig:necklace}
\end{figure}

We first elaborate on our design of the wearable component. We use a Knowles BU-27135 
miniature \acc with the dimension of only 7.92$\times$5.59$\times$2.28mm so that it can 
easily fit in any wearable design. The \acc uses only the z-axis and has an analog bandwidth of 11kHz, 
enough to capture the bandwidth of a speech signal. We utilize an external Bluetooth 
transmitter that provides Analog-to-Digital Conversion (ADC) and Bluetooth 
transmission capabilities to the voice assistant extension. To reduce 
energy consumption, \name starts streaming the \acc signal only upon 
request from the voice assistant. 
Our prototype communicates the \mic and \acc signals to a Matlab-based server which 
performs the matching and returns the result to the voice assistant. 
Fig.~\ref{fig:necklace} depicts our wireless prototype standalone, and 
attached to a pair of eyeglasses.

Our system is integrated with Google Now voice assistant to enable voice command authentication. 
\name starts execution immediately after the start of a voice session (right after ``OK Google'' is 
recognized). It blocks the voice assistant's command execution after the voice 
session ends until the matching result becomes available. If the matching fails, \name
kills the voice session. To achieve its functionality, \name intercepts
both the \texttt{HotwordDetector} and the \texttt{QueryEngine} to establish the required 
control flow.

Our voice assistant extension is implemented as a standalone user-level service. It is responsible 
for retrieving \acc signals from the wearable device, and sending both \acc and \mic to our Matlab-based
server for analysis. The user-level service provides two RPC methods, \texttt{start} and \texttt{end}, 
which are triggered by the events generated when the hotword ``OK Google'' is detected, and 
when the query (command) gets executed, respectively. The first event can be observed by filtering 
the Android system logs, and we intercept the second by overriding the Android IPC mechanisms, by 
filtering the Intents sent by Google Now. 
Also, since some Android devices (e.g., Nexus 5) do not allow two apps to access 
the microphone at the same time, we need to stream the voice signal retrieved by the voice 
assistant to our user-level service. We solve this by intercepting the \texttt{read} method in 
the \texttt{AudioRecord} class. Whenever Google Now gets the updated voice data through this 
interface, it will forward a copy of the data to our user-level service via another RPC method.

Note that the modifications and interceptions above are necessary only because we 
have no access to the Google Now source. The incorporation of \name is straightforward in the 
cases when developers try to build/extend their voice assistant.

\subsection{Usability}
\label{sec:usability}

\begin{figure}[t]
     \centering
     \subfigure[Earbuds]{\includegraphics[width=0.3\columnwidth]{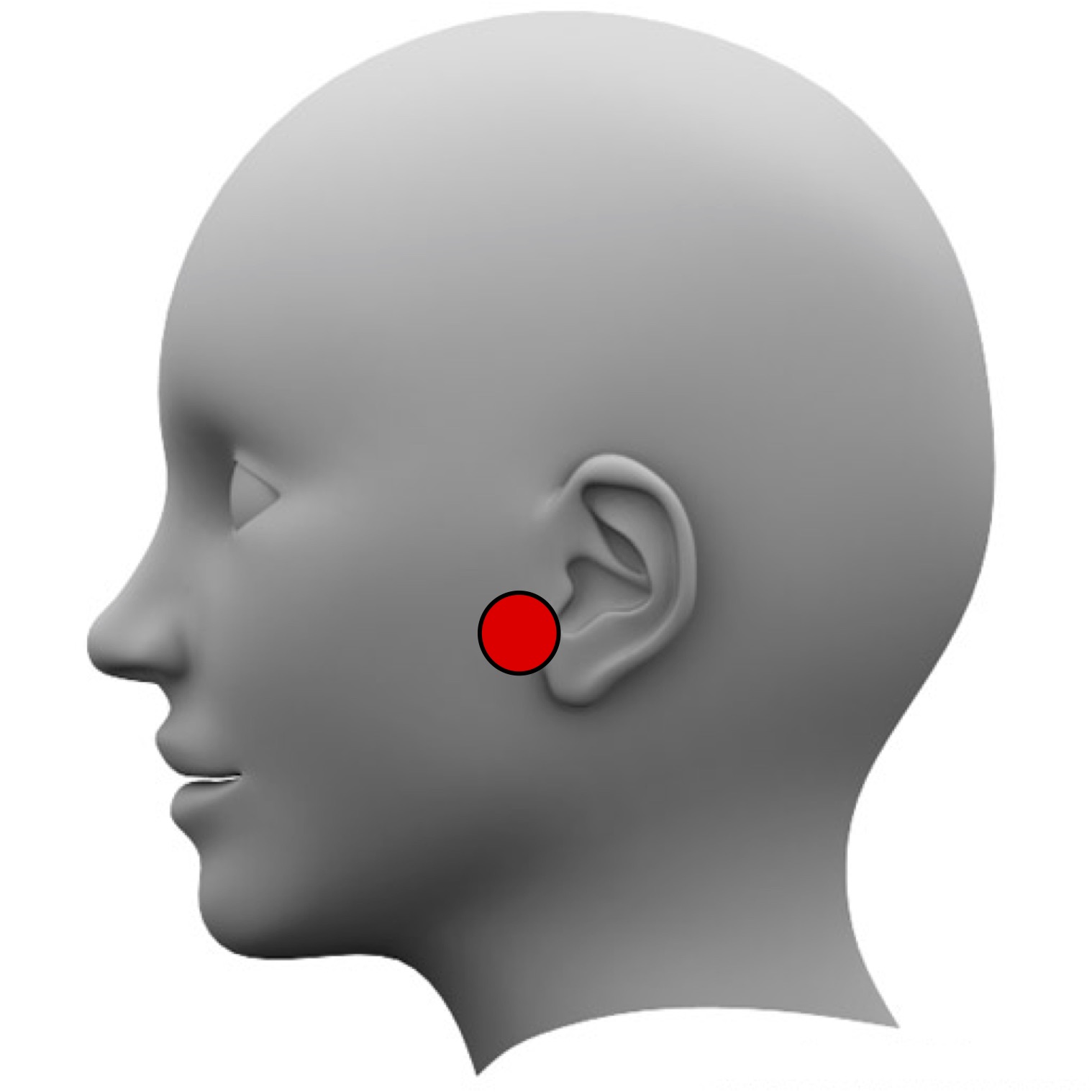}}
     \hfill  
     \subfigure[Eyeglasses]{\includegraphics[width=0.3\columnwidth]{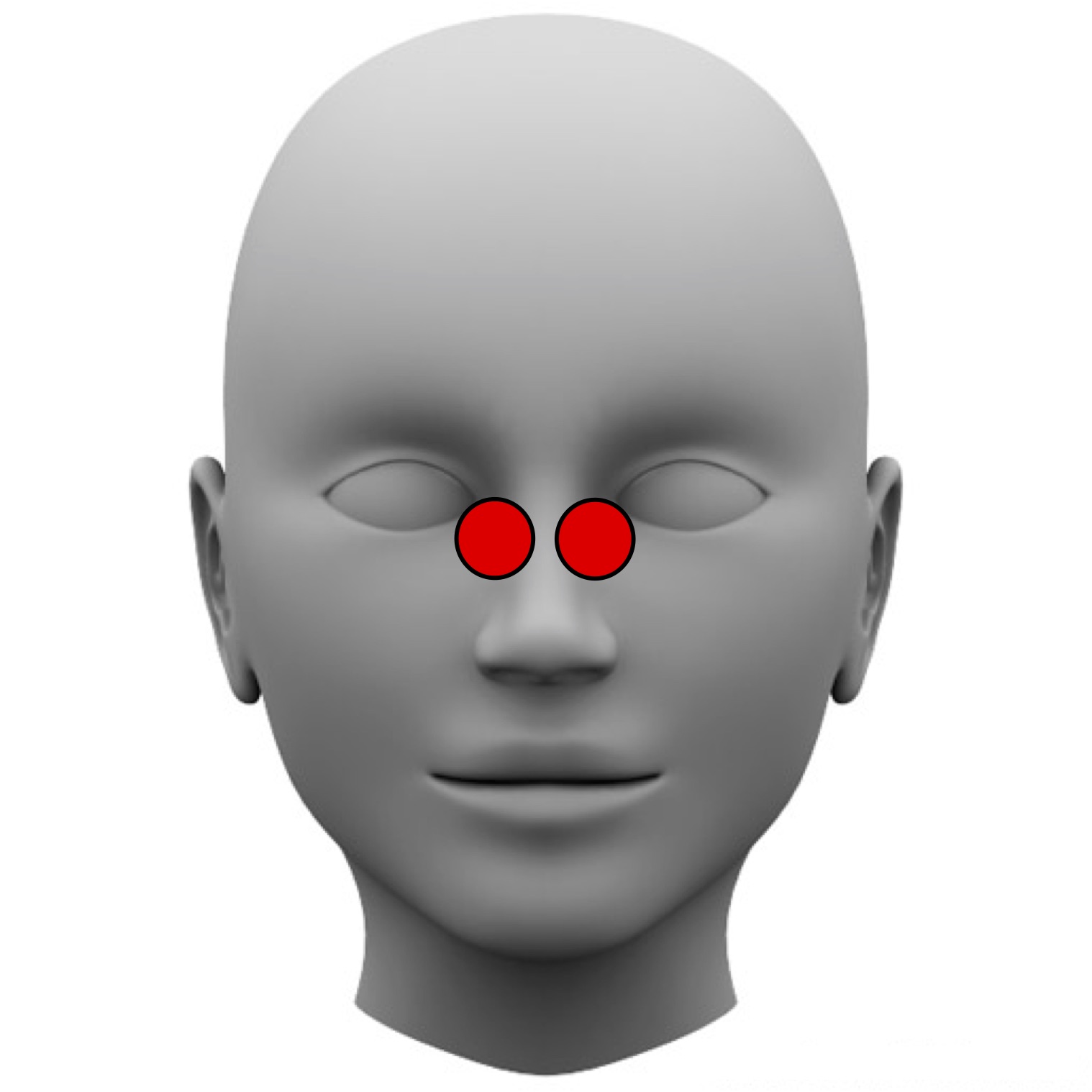}}
     \hfill
     \subfigure[Necklace]{\includegraphics[width=0.3\columnwidth]{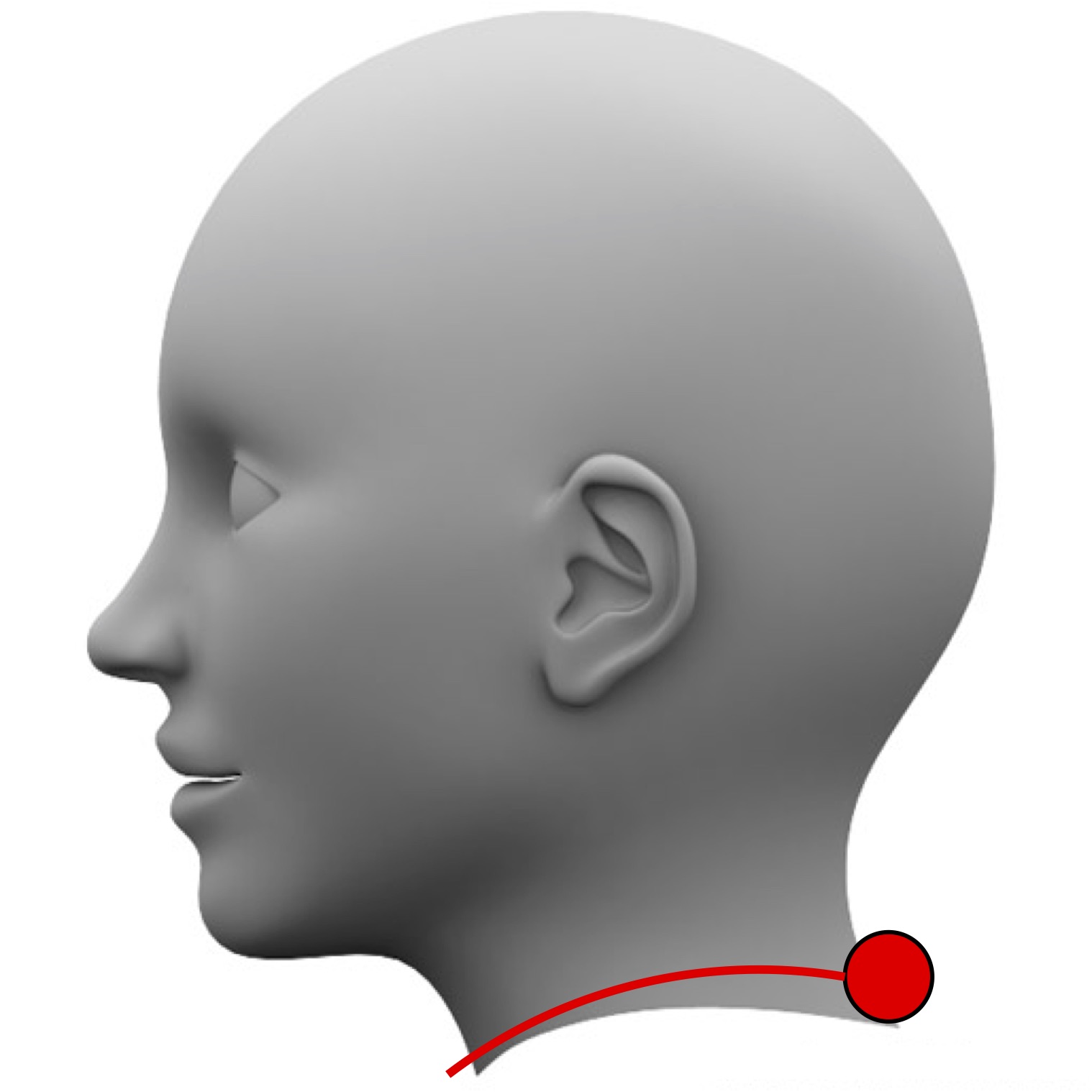}}
     \hfill     
     \caption{The wearable scenarios supported by \name.}
     \label{fig:scenarios}
\end{figure}

\name requires the user to wear a security-assisting device. There are two general ways 
to meet this requirement. The first is to ask users to wear an additional 
device for security, while the other is to embed \name in existing wearable products that the 
users are already comfortable with in their daily lives. We opted for the latter
as security has always been a secondary concern for users~\cite{West:2008}. 
Our prototype supports three widely-adopted wearable scenarios: earbuds/earphones/headsets, 
eyeglasses, and necklace/lockets. Fig.~\ref{fig:scenarios} shows the 
positions of the \acc in each scenario. We select 
these areas because they have consistent contact with the user's body.
While \name performs well on all facial areas, shoulders and the sternal 
surface, we only focus on the three positions shown in Fig.~\ref{fig:scenarios} since they 
conform with widely-adopted wearables.

We have conducted a usability survey to study the users' acceptance of the different 
configurations of \name. We surveyed 952 individuals using Amazon Mechanical Turk. 
We restricted the respondent pool to those from the US with previous experience with 
voice assistants. We compensated each respondent with \$0.5 for their participation. 
Of the respondents, 40\% are female, 60\% are employed full-time, and 67\% 
have an education level of associate degree or above. Our respondents primarily use voice 
assistants for information search (70\%), navigation (54\%), and communication (47\%). 
More than half (58\%) of them reported using a voice assistant at least once a week. 

\textbf{Survey Design:}
We follow the USE questionnaire methodology~\cite{lund:2001} to measure the usability 
aspects of \name. We use a 7-point Likert scale (ranging from Strongly Disagree to 
Strongly Agree) to assess the user's satisfaction with a certain aspect or configuration 
of \name.  We pose the questions in the form of how much the respondent agrees with 
a certain statement, such as: \textit{I am willing to wear a necklace that contains 
the voice assistant securing technology.} Below, we report a favorable result as the 
portion of respondents who answered a question with a score higher than 4 (5,6,7) on 
the 7-point scale. Next to each result, we report the portion of those surveyed, between 
brackets, who answered the question with a score higher than 5 (6 or 7). 

The survey consists of three main parts that include: demographics and experience
with voice assistants, awareness of the security issues, and the perception towards \name. 
In Section~\ref{sec:evaluation}, we will report more on the other usability aspects of 
\name, such as matching accuracy, energy, and latency.

\textbf{Security Awareness:}
We first asked the respondents about their opinion regarding the security of voice 
assistants. Initially, 86\% (63\%) of the respondents indicate that they think
the voice assistants are secure. We then primed the respondents about the security 
risks associated with voice assistants by iterating the attacks presented in 
Section~\ref{sec:system-threat-models}. Our purpose was to study the perception 
of using \name from individuals who are already aware of the security problems of voice assistants. 

After the priming, the respondents' perceptions shifted considerably. 
71\% (51\%) of the respondents indicate that attacks to voice assistants are dangerous, 
and 75\%(52\%) specified that they would take steps to mitigate the threats. 
Almost all of the latter belong to the set of respondents who now regard these attacks 
as dangerous to them.
 
 \begin{figure}
	\centering
	\includegraphics[width=0.8\columnwidth]{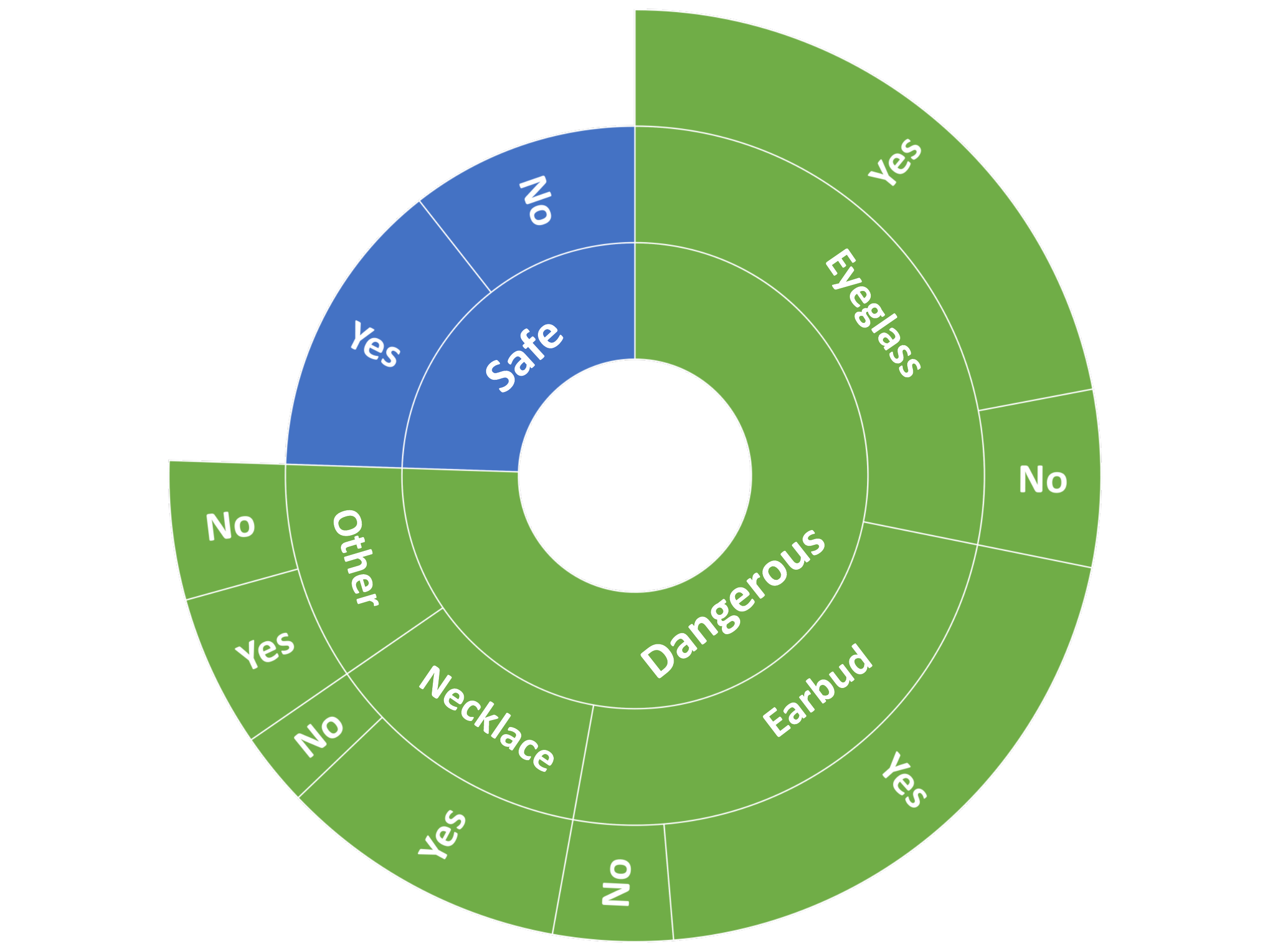}
	\caption{A breakdown of respondents' wearability preference by security concern and 
     daily wearables. \textit{Dangerous} and \textit{Safe} refer to participants' attitudes 
     towards the attacks to voice assistants after they've been informed; the \textit{Dangerous} 
     category is further split according to the wearables that people are already wearing on a 
     daily basis; \textit{Yes} and \textit{No} refer to whether participants are willing to use 
     \name in at least one of three settings we provided.}
	\label{fig:usability_analysis}
\end{figure}

\textbf{Wearability:}
In the last part of the survey, we ask the participants about their 
preferences for wearing \name in any of the three configurations of 
Fig.~\ref{fig:scenarios}. We have the following takeaways from the 
analysis of survey responses.
\begin{itemize}
\item 70\%(47\%) of the participants are willing to wear at least one of 
\name's configurations to provide security protection. These respondents are 
the majority of those who are strongly concerned about the security threats.

\item 48\% (29\%) of the respondents favored the earbuds/earphone/headset option, 
38\% (23\%) favored the eyeglasses option and 35\% (19\%) favored the necklace/locket 
option. As expected, the findings fit the respondents' wearables in their daily 
lives. 71\% of the respondents who wear earbuds on a daily basis favored that 
option for \name, 60\% for eyeglasses and 63\% for the necklace option.

\item There is no discrepancy in the wearable options among both genders. The gender
distribution of each wearable option followed the same gender distribution of the whole
respondent set.
\item More than 75\% of the users are willing to pay \$10 more for a wearable 
equipped with this technology while more than half are willing to pay \$25 more.
 
\item Respondents were concerned about the battery life of \name. 
A majority of 73\% (81\%) can accommodate charging once a week, 
60\% (75\%) can accommodate once per 5 days, and 38\% (58\%) can accommodate 
once each three days. In Section~\ref{sec:evaluation}, we show that the 
energy consumption of \name matches the respondents' requirements.
\end{itemize}

Fig.~\ref{fig:usability_analysis} presents a breakdown of the major findings in 
our usability survey. These results demonstrate that users are willing to 
accept the different configurations of \name, especially when they are concerned 
about the privacy/security threats and when \name comes in the forms of which they are 
already comfortable with.

\section{Matching Algorithm}
\label{sec:matching}

\begin{figure}[t]
     \centering
     \subfigure[Raw input signals]{\includegraphics[width=0.49\columnwidth]{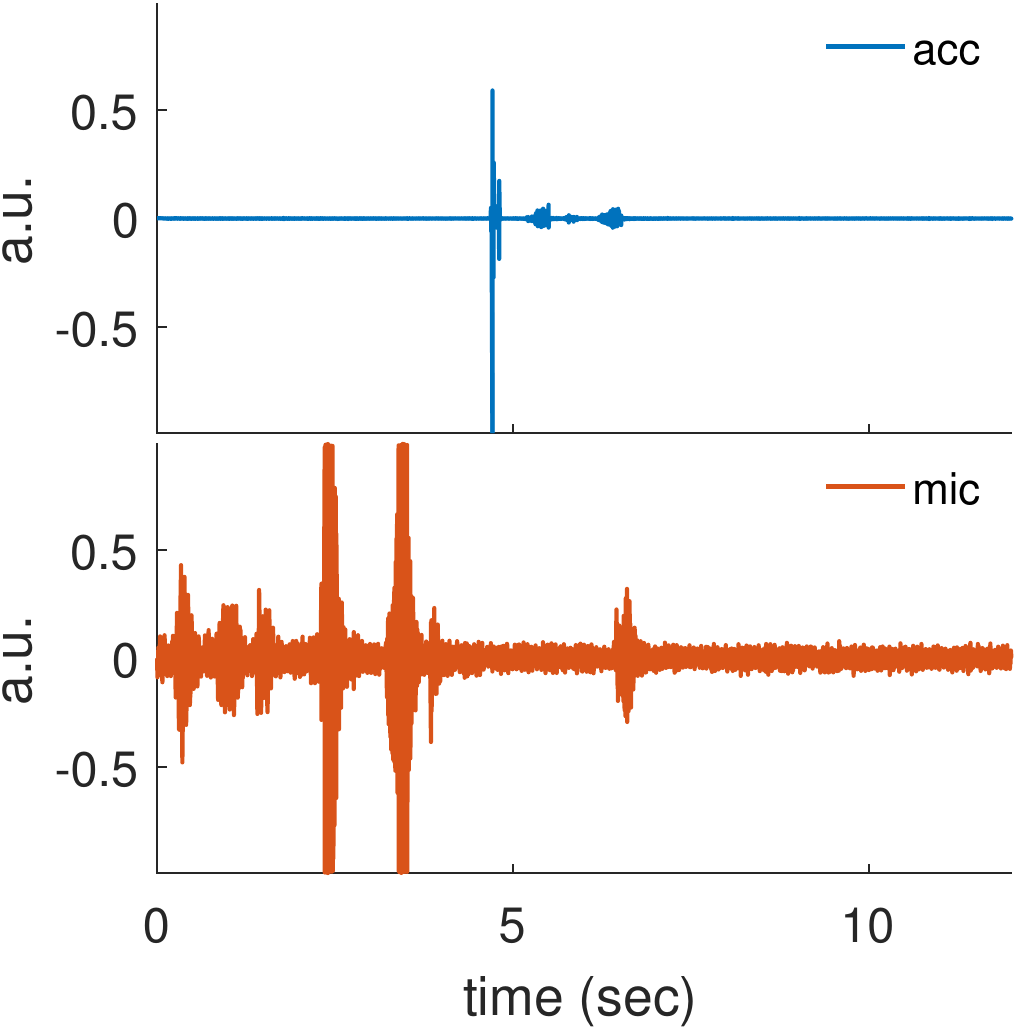}\label{sig-raw}}
     \hfill  
     \subfigure[Energy envelopes]{\includegraphics[width=0.49\columnwidth]{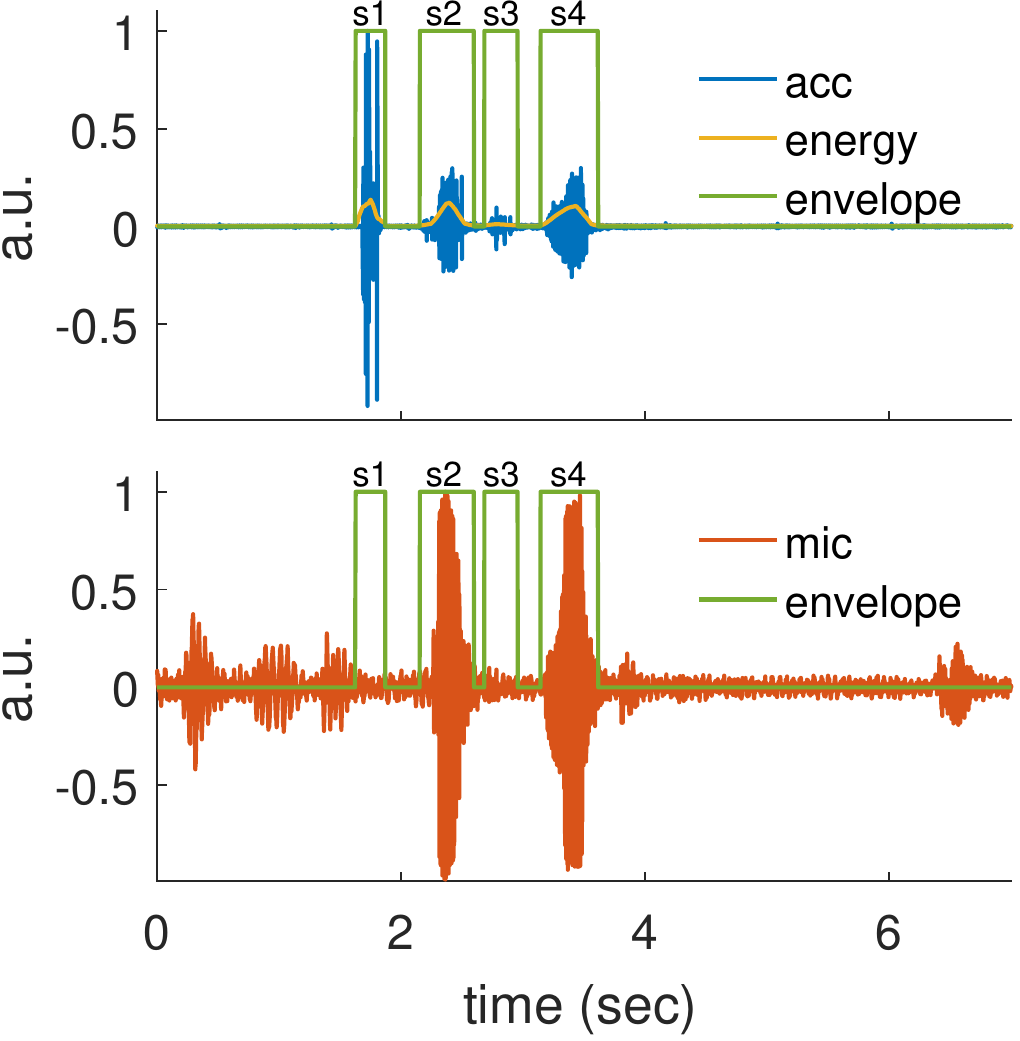}\label{sig-energy}}
     \hfill
     \caption{Pre-processing stage of \name's matching.}
     \label{fig:match_pre}
\end{figure}

The matching algorithm of \name (highlighted in Fig.~\ref{fig:arch}) 
takes as input the speech and vibration signals along with 
their corresponding sampling frequencies. It outputs a decision value indicating whether 
there is a match between the two signals as well as a ``cleaned" speech signal
in case of a match. \name performs the matching in three stages:
{\em pre-processing}, {\em speech segments analysis}, and {\em matching decision}.

In what follows, we elaborate on \name's matching algorithm using a running example of
a male speaker recording the two words: ``cup" and ``luck" with a short pause between them. 
The speech signal is sampled by an \acc from the lowest point on the sternum at 64kHz and 
recorded from a built-in laptop \mic at a sampling frequency of 44.1kHz, 50cm away from the speaker.

\subsection{Pre-processing}

First, \name applies a highpass filter, with cutoff frequency at 100 Hz,
to the \acc signal. The filter removes all the artifacts of the low-frequency user movement
to the \acc signal (such as walking or breathing). We use 100Hz as a cutoff threshold
because humans cannot generate more than 100 mechanical movements per second.
 \name then 
re-samples both \acc and \mic signals to the same sampling rate while applying 
a low-pass filter at 4kHz to prevent aliasing.
We choose a sampling rate of 8kHz that preserves most acoustic features of the speech signal 
and reduces the processing load. Thus, \name requires an \acc of bandwidth larger than 4kHz. 
Then \name applies 
Fig.~\ref{sig-raw} shows both raw signals immediately after both signals are filtered and resampled. 
As evident from the figure, the \acc signal has a high-energy spike due to the sudden movement 
of the \acc (e.g., rubbing against the skin), and small energy components resulting from speech vibrations. 
On the other hand, the speech signal has two high-energy segments 
along with other lower-energy segments corresponding to background noise.

Second, \name normalizes the magnitude of both signals to have a maximum magnitude of unity, 
which necessitates removal of the spikes in the signals. Otherwise, the lower-energy components
referring to the actual speech will not be recovered. The matching algorithm computes
a running average of the signal's energy and enforces a cut-off threshold, keeping only the 
signals with energy level within the moving average plus six standard deviation levels.

After normalizing the signal magnitude, as shown in the top plot of Fig.~\ref{sig-energy},
\name aligns both signals by finding the time shift that results in the maximum cross correlation
of both signals. Then, it truncates both signals to make them have the same length.
Note that \name does not utilize more sophisticated alignment algorithms such as Dynamic
Time Warping (DTW), since they remove timing information critical to the signal's pitch and
they also require a higher processing load. Fig.~\ref{sig-energy} shows both \acc and
\mic signals aligned and normalized.

The next pre-processing step includes identification of the energy envelope
of the \acc signal and then its application to the \mic signal.
\name identifies the parts of the signal that have a significant signal-to-noise ratio (SNR). 
These are the ``bumps" of the signal's energy as shown in the top plot of Fig.~\ref{sig-energy}.
The energy envelope of the signal is a quantification of the signal's energy
between 0 and 1. In particular, the portions of the signal with average energy 
exceeding 5\% of maximum signal energy map to 1, and other segments map to 0. 
This results in four energy segments of the \acc signal of Fig.~\ref{sig-energy}.
The thresholds for energy detection depend on the average noise level
(due to ADC chip's sampling and quantization)
when the user is silent. We chose these thresholds after studying our wireless
prototype's Bluetooth transmitter. 

Finally, \name applies the \acc envelope to the \mic signal so that it removes all parts from the \mic signal
that did not result from body vibrations, as shown in the bottom plot of Fig.~\ref{sig-energy}.
This is the first real step towards providing the security guarantees. In most cases, it avoids attacks 
on voice assistant systems when the user is not actively speaking. Inadvertently, it improves the 
accuracy of the voice recognition by removing background noise and sounds from the speech 
signals that could not have been generated by the user.  
   
 \begin{figure}[t]
     \centering
     \subfigure[Non-matching segment \textit{s1}]{\includegraphics[width=0.48\columnwidth]{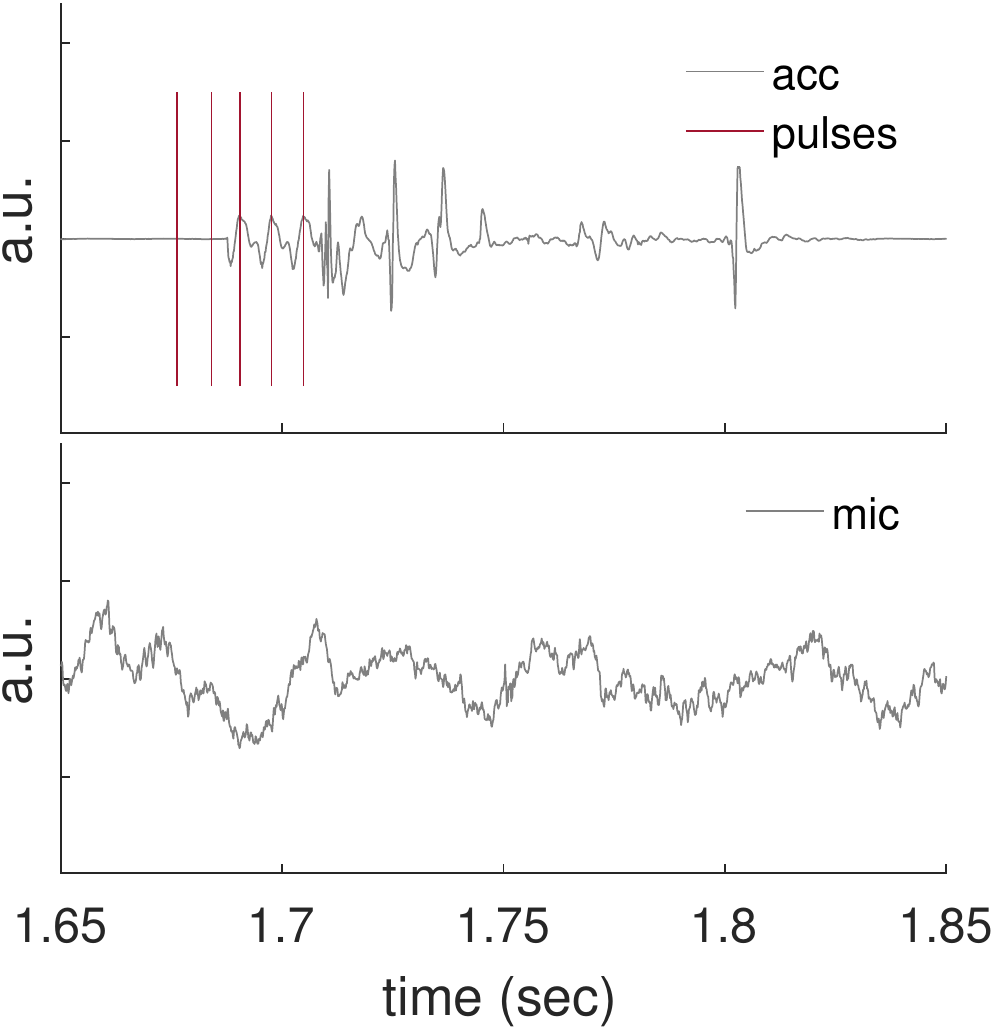}
     \label{sig-glottal-no-match}}
     \hfill  
     \subfigure[matching segment \textit{s4}]{\includegraphics[width=0.48\columnwidth]{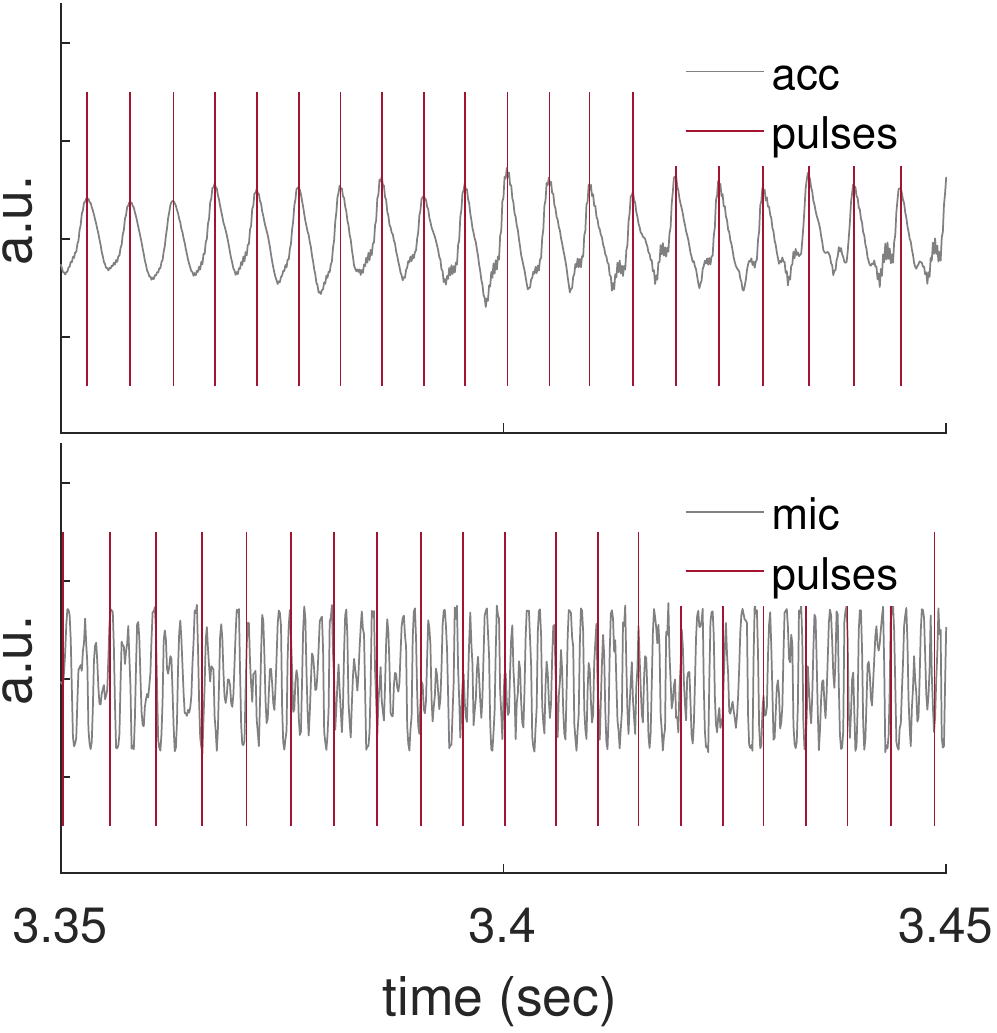}
     \label{sig-glottal-match}}
     \caption{Per-segment analysis stage of \name.}
     \label{fig:match_seg}
\end{figure}

\subsection{Per-Segment Analysis}

Once it identifies high-energy segments of the \acc signal, \name starts a 
segment-by-segment matching. Fig.~\ref{sig-energy} shows four segments corresponding
to the parts of the signal where the envelope is equal to 1. 

For each segment, \name normalizes the signal magnitude to unity to remove the effect of 
other segments, such as the effect of the segment \textit{s1} in Fig.~\ref{sig-energy}.
This serves to make the energy content of each segment uniform, which will elaborate 
on later in Section~\ref{sec:model}.
\name then applies the approach of Boersma~\cite{Boersma:1993} to extract the glottal cycles from each segment. 
The approach relies on the identification of periodic patterns in the signal as the local maxima of the 
auto-correlation function of the signal. Thus, each segment is associated with a series of 
glottal pulses as shown in Fig.~\ref{fig:match_seg}.
\name uses information about the segment and the corresponding glottal pulses 
to filter out the segments that do not correspond to human speech and 
those that do not match between the \acc and \mic signals as follows.

\begin{enumerate}

\item If the length of the segment is less than 20ms, the length of a single phoneme, then \name
removes the segment from both \acc and \mic signals. Such segments might arise from sudden noise.

\item If the segment has no identifiable glottal pulses or the length of the longest continuous 
sequence of glottal pulses is less than 20ms (the duration to pronounce a single phoneme),
then \name also removes the segment. 
Fig.~\ref{sig-glottal-no-match} shows the segment ``s1" at a higher resolution. 
It only contains five pulses which could not have resulted from a speech. 

\item If the average glottal cycle of the \acc segment is larger than 0.003sec or smaller than
0.0125sec, then \name removes the segment from both signals. This refers to the case of the fundamental 
frequency falling outside the range of [80Hz, 333 Hz] which corresponds to the human speech range.

\item If the average relative distance between glottal pulse sequence between the \acc and \mic segments is 
higher than 25\%, then \name removes the segment from both signals. This refers to the case of interfered
speech (e.g., attacker trying to inject speech);
the instantaneous pitch variations should be similar between the \acc and \mic~\cite{Mehta:2016} 
in the absence of external interference. For example, it is evident that the pitch information is very different
between the \acc and \mic of Fig.~\ref{sig-glottal-no-match}. 

\end{enumerate}

After performing all the above filtering steps, \name does a final verification step by running a normalized 
cross correlation between the \acc and \mic segments. If the maximum correlation coefficient falls inside 
the range [-0.25,0.25], then the segments are discarded. We use this range as a conservative way
of specifying that the segments do not match (correlation coefficient close to zero).
The correlation is a costlier operation but is a known 
metric for signal similarity that takes into consideration all the information of the time-domain signals. 
For example, the segment ``s4" depicted in Fig.~\ref{sig-glottal-match} shows matching pitch information 
and a maximum cross-correlation coefficient of 0.52.

\begin{figure}[t]
     \centering
     \subfigure[][Output signals]{\includegraphics[width=0.49\columnwidth]{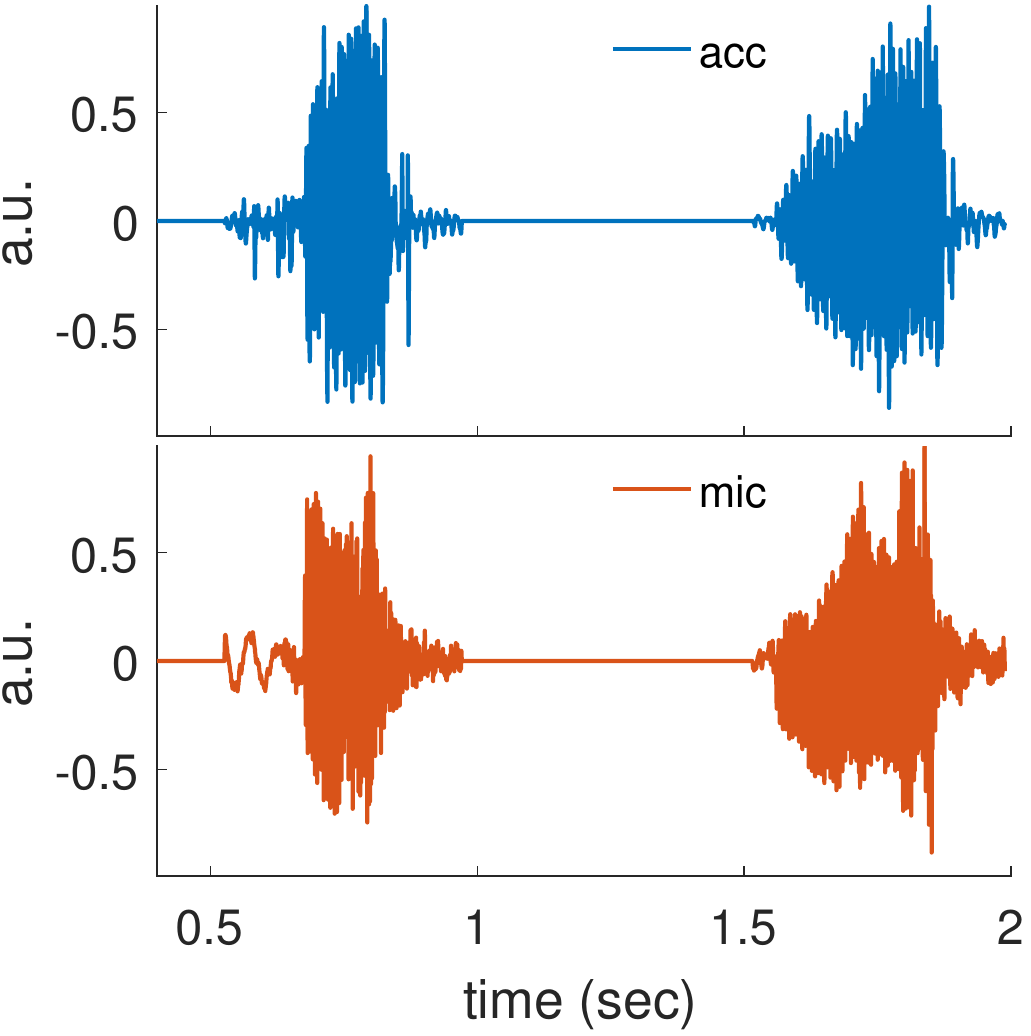}\label{sig-post}}
     \hfill  
     \subfigure[][Cross correlation]{\includegraphics[width=0.49\columnwidth]{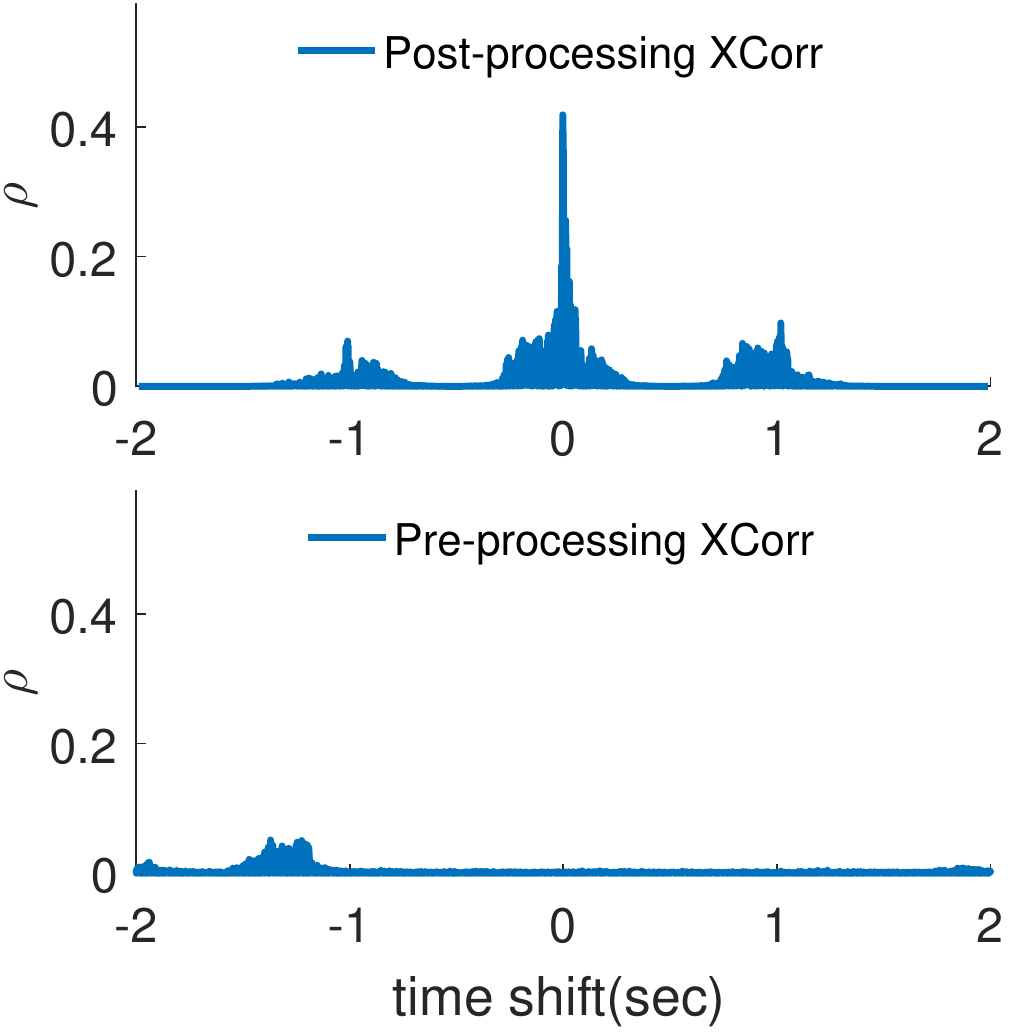}\label{sig-corr}}
     \hfill
     \caption{Matching decision stage of \name's matching.}
     \label{fig:match_post}
\end{figure}

\subsection{Matching Decision}
\label{sec:matching_decision}
After the segment-based analysis finishes, only the ``surviving" segments comprise 
the final \acc and \mic signals. In Fig.~\ref{sig-post}, only the segments ``s2" and 
``s4" correspond to matching speech components. It is evident from the bottom plot 
that the \mic signal has two significant components referring to each word. 

The final step is to produce a matching decision. \name measures the similarity 
between the two signals by using the normalized cross-correlation,
as shown in the top plot of Fig.~\ref{sig-corr}. 
\name cannot just perform the cross-correlation on the input signals before cleaning. 
Before cleaning the signal, the cross-correlation results do not have any real indication of signal similarity. 
Consider the lower plot of Fig.~\ref{sig-corr}, which corresponds to the cross-correlation performed 
on the original input signals of Fig.~\ref{sig-raw}. As evident from the plot, the cross-correlation
shows absolutely no similarity between the two signals, even though they
describe the same speech sample.

Instead of manually constructing rules that map the cross-correlation
vector to a \textit{matching} or \textit{non-matching} decision, we opted to
utilize a machine learning-based classifier to increase the accuracy of \name's 
matching.
Below, we elaborate
on the three components of \name's classifier: the feature set, the machine
learning algorithm and the training set.

\paragraph*{\textbf{Feature Set}}
In general, the feature vector comprises the normalized cross-correlation
values ($h(t)$) of the final \acc and \mic signals. However, we need
to ensure that the structure of the feature vector is uniform
across all matching tasks. To populate the feature vector,
we identify the maximum value of $h(t)$, and then uniformly sample 500 points to the left 
and another 500 to the right of the maximum. We end up with a feature vector containing 
1001 values, centered at the maximum value of the normalized cross-correlation.

Formally, if the length of $h(t)$ is $t_e$, let $t_m = \underset{t}{\operatorname{arg\,max}}{\abs{h(t)}}$.
Then, the left part of the feature vector is $h_l[n]=h(\frac{n.t_m}{500}),~ 1<n<500$.
The  right part of the feature vector is $h_r[n]=h(t_m + \frac{n.(t_e-t_m)}{500}), 1<n<500$. 
The final feature vector can then be given 
as $h[n] = h_l[n] + h(t_m).\delta[n-501] + h_r[n-502].$

\paragraph*{\textbf{Classifier}}
We opted to use SVM as the classifier thanks to its ability to
deduce linear relations between the cross-correlation values
that define the feature vector. We utilize Weka~\cite{weka} to train
an SVM using the Sequential Minimal Optimization (SMO) algorithm~\cite{smo}. 
The SMO algorithm uses a logistic calibrator with neither
standardization nor normalization to train the SVM. The SVM
utilizes a polynomial kernel with the degree equal to 1. We use
the trained model in our prototype to perform the online classification.

\paragraph*{\textbf{Training Set}}
Here, it is critical to specify that our training set has been 
generated offline and is user-agnostic; we performed the training only once. 
We recorded (more on that in Section~\ref{sec:phonetics})
all 44 English phonemes (24 vowels and 20 consonants)
from one of the authors at the lower sternum position
using both the \acc and \mic. Hence, we have 44 \acc ($acc(i)$) and \mic ($mic(i)$) pair
of recordings corresponding for each English phoneme. To generate the training set,
we ran \name's matching over all 44 $\cross$ 44 \acc and \mic recordings
to generate 1936 initial feature vectors, ($fv$), and their labels as follows:
\begin{align*}
\forall i: 1\le i \le 44;\quad \forall j: 1 \le j \le 44; \\
fv[j+44(i-1)] = match(acc(i),mic(j))\\
label[j+44(i-1)] = 1_{i=j}.
\end{align*}

The generated training dataset contains only 44 vectors with positive labels.
This might bias the training process towards the majority class ($label=0$). To counter this
effect, we amplified the minority class by replicating the vectors with positive
labels five times. The final training set contains 236 vectors with positive labels
and 1892 vectors with negative labels. We use this training set to train
the SVM, which, in turn, performs the online classification.  

\name's classifier is trained \textit{offline},
only \textit{once} and \textit{only} using a single training set.
The classifier is thus agnostic of the user, position on the body and language. In our
user study and rest of the evaluation of Section~\ref{sec:evaluation},
this (same) classifier is used to perform all the matching. To use \name,
the user {\em need not} perform any initial training.

After computing the matching result, \name passes the final (cleaned and normalized) \mic signal 
to the voice assistant system to execute the speech recognition and other functionality.

\section{Phonetic-Level Analysis}
\label{sec:phonetics}

We evaluate the effectiveness of our matching algorithm on 
phonetic-level matchings/authentications. The International 
Phonetic Alphabet (IPA) standardizes the representation
of sounds of oral languages based on the Latin alphabet. While the number of words in 
a language, and therefore the sentences, can be uncountable, the number of phonemes 
in the English language are limited to 44 vowels and consonants. 
By definition, any English word or sentence, as spoken by a human, is necessarily 
a combination of those phonemes~\cite{phonetic};
A phoneme\footnote{https://www.google.com/search?q=define:phonemes}
represents the smallest unit of perceptual sound. Our phonetic-level
evaluation represents a baseline of \name's operation.
Table~\ref{table:phonetics} of 
Appendix A lists 20 vowels and 24 consonants phonemes, with two words representing 
examples of where the phonemes appear.

We study if \name can correctly match the English phoneme 
between the \acc and \mic (true positives),
and whether it mistakenly matches phoneme samples from 
\acc to other phoneme samples from the \mic (false positives).

We recruited two speakers, a male and a female, to record the 44 examples
listed in Table~\ref{table:phonetics}. Each example comprises two
words, separated by a brief pause, both representing a particular phoneme.
We asked the speaker to say both words, not just the phoneme, as it is
easier for the speaker to pronounce the phoneme in the context of a word.
Both participants were 50cm away from the built-in microphone
of an HP workstation laptop. At the same time, both speakers
were wearing \name, with the \acc taped to the sternum. 
The \mic was sampled at 44100 Hz, and the \acc at 64000 Hz.

\subsection{Accelerometer Energy \& Recognition}
\label{sec:acc_energy}

\begin{figure}[t]
     \centering
     \subfigure[Relative energy of \acc signal to \mic signal]{\includegraphics[width=0.47\columnwidth]{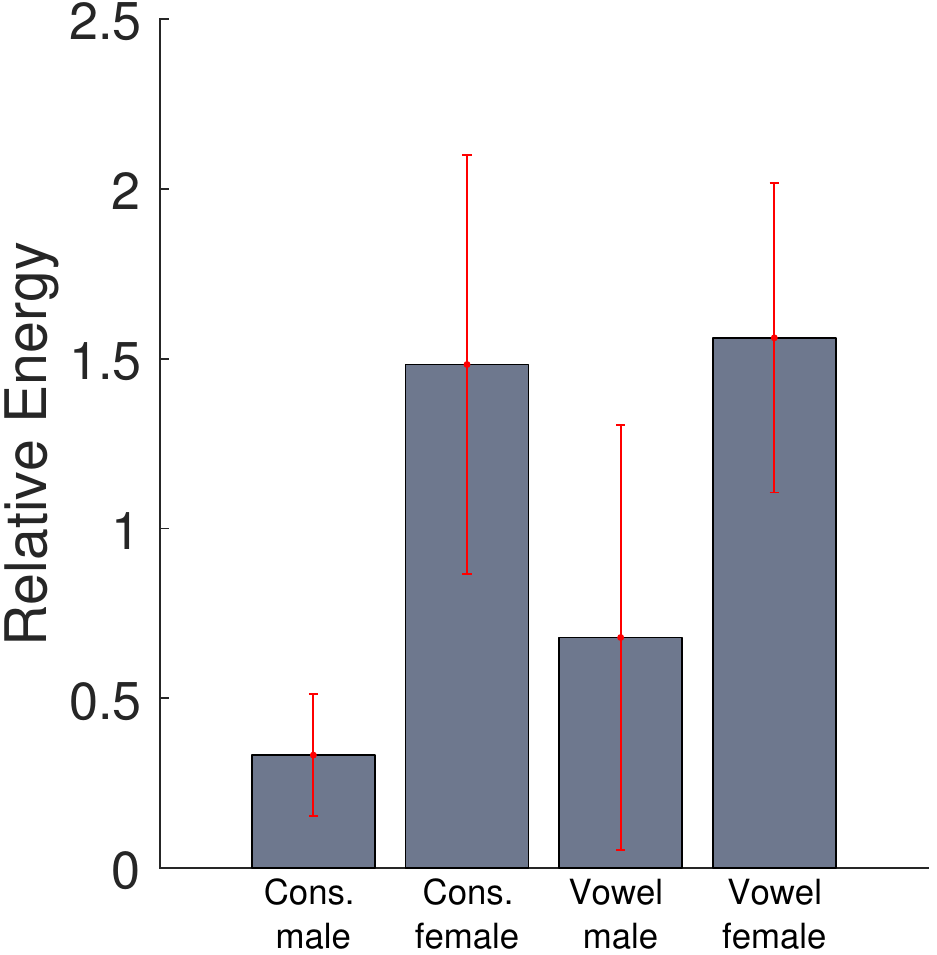}\label{fig:relative-energy}}
     \hfill  
     \subfigure[ASR accuracy for \acc signal]{\includegraphics[width=0.47\columnwidth]{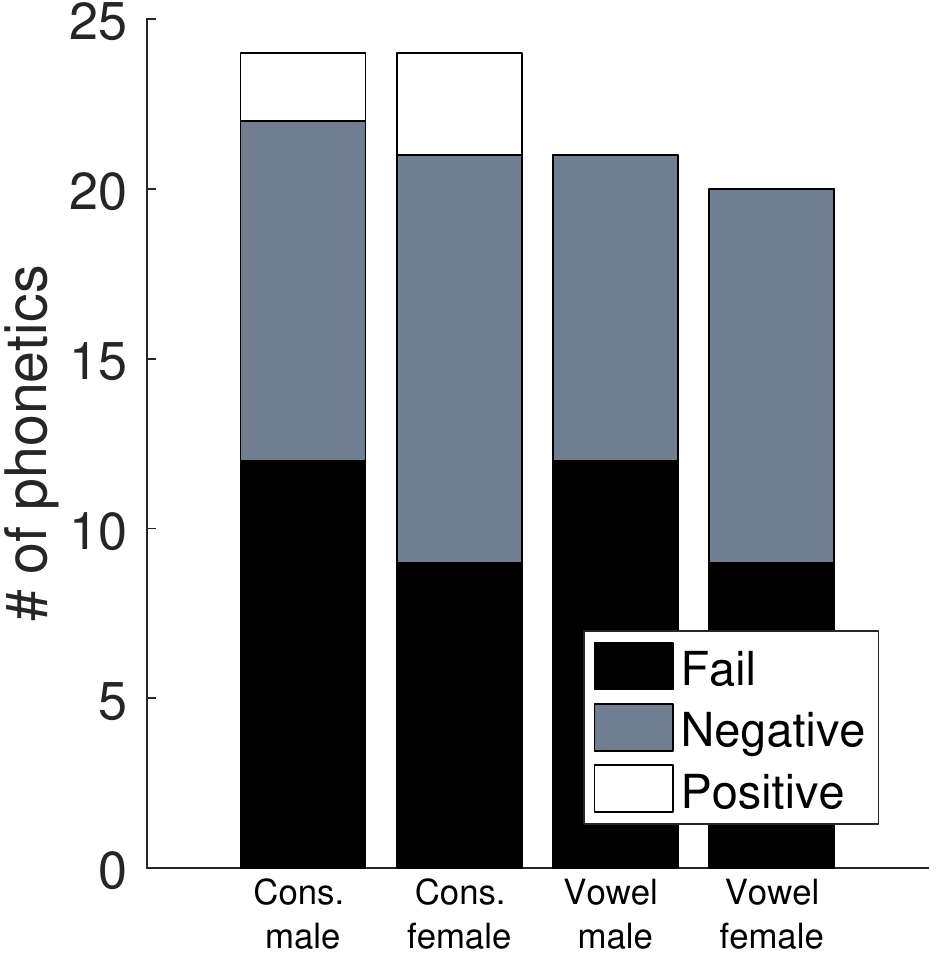}\label{fig:asr-acc}}
     \hfill
     \caption{Analysis of the vibrations received by the \acc.}
     \label{fig:acc-energy}
\end{figure}

Phonemes originate from a possibly different part of the chest-mouth-nasal
area. In what follows, we show that each phoneme results in vibrations that 
the \acc chip of \name can register, but 
does not retain enough acoustic features to substitute
a \mic speech signal for the purpose of voice recognition. 
This explains our rationale for employing the matching-based approach.

We perform the pre-processing stage of \name's matching algorithm to clean
both \acc and \mic signals for each phoneme. After normalizing
both signals to a unity magnitude, we compute the \acc signal's
energy relative to that of the \mic. Fig.~\ref{fig:relative-energy} depicts
the average relative energy of the vowel and consonants phonemes 
for both the female and male speakers.

\textit{All}
phonemes register vibrations, with the minimum relative energy (14\%) coming from the 
\ipa{\:OI} (the pronunciation of ``oy" in ``boy") phoneme of the male speaker. It is also clear from 
the figure that there is a low discrepancy of average relative energy between
vowels and consonants for the same speaker. Nevertheless, we notice
a higher discrepancy between the two speakers for the same phoneme. The female
speaker has a shorter distance between the larynx and lowest point of the sternum,
and she has a lower body fat ratio so that the chest skin is closer 
to the sternum bone. It is worth noting that the energy reported in the figure
does not represent the energy at the time of the recording but after the initial
processing and normalization. This explains why in some cases the \acc energy exceeds
that of the \mic.

While the \acc chip senses considerable energy from the chest vibrations, it 
cannot substitute for the \mic. To confirm this, we passed the recorded and 
cleaned \acc samples of all phonemes for both speakers to the Nuance Automatic
Speaker Recognition (ASR) API~\cite{nuance:api}. 
Fig.~\ref{fig:asr-acc} shows the breakdown of voice recognition accuracy for the \acc
samples by phoneme type and speaker. Clearly, a state-of-the-art
ASR engine fails to identify the actual spoken words. In particular,
for about half of the phonemes for both speakers, the ASR fails to return any result. 
Nuance API returns three suggestions for each \acc sample for the rest of the phonemes. 
These results do not match any of the spoken words.
In only three cases for consonants phonemes for both speakers, the API
returns a result that matches at least one of the spoken words.

The above indicates that existing ASR engines cannot interpret
the often low-fidelity \acc samples, but it does not indicate that
ASR engines cannot be retrofitted to recognize samples with higher accuracy.
This will, however, require significant changes to deploying and training
these systems. On the other hand, \name is an entirely
client-side solution that requires no changes to the ASR engine
or the voice assistant system.

\begin{table}[t]

\renewcommand{\arraystretch}{1.1}
\footnotesize
  \centering
   \caption{The detection accuracy of \name for the English phonemes.}
\begin{tabular}{L{2cm} L{2cm} C{1.2cm} C{1.2cm} }
\toprule
\textbf{\mic} & \textbf{\acc} & \textbf{TP (\%)} & \textbf{FP (\%)} \\ 

\midrule 
consonants & consonants & 90 & 0.2 \\
consonants & vowels & - & 1.0 \\
vowels & consonants & - & 0.2 \\
vowels & vowels & 100 & 1.7 \\
\midrule

all & all & 94 & 0.7 \\

\bottomrule
\end{tabular} 

\label{table:phonetic_eval}
\end{table}

\begin{figure}[t]
     \centering
     \subfigure[Low energy and white noise]{\includegraphics[width=0.43\columnwidth]{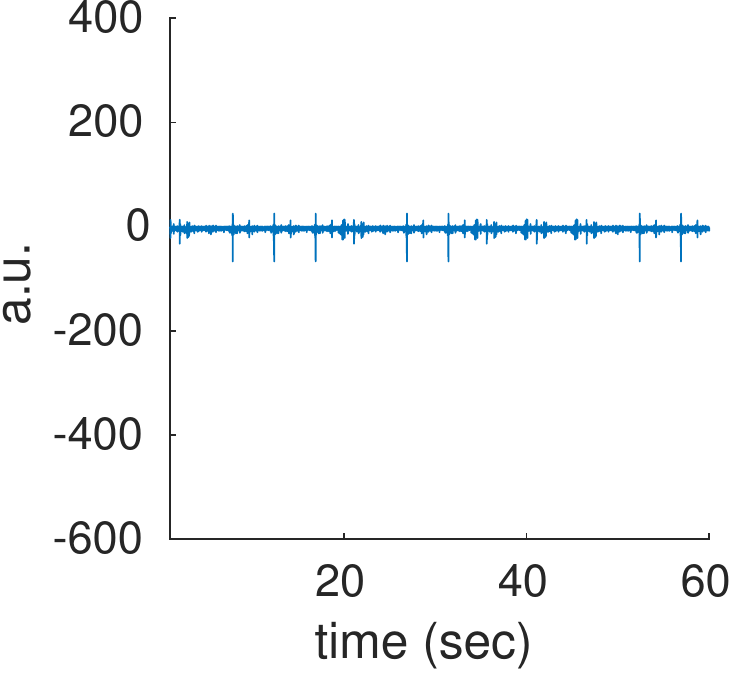}\label{fig:low-noise}}
     \hfill  
     \subfigure[High energy and periodic noise]{\includegraphics[width=0.43\columnwidth]{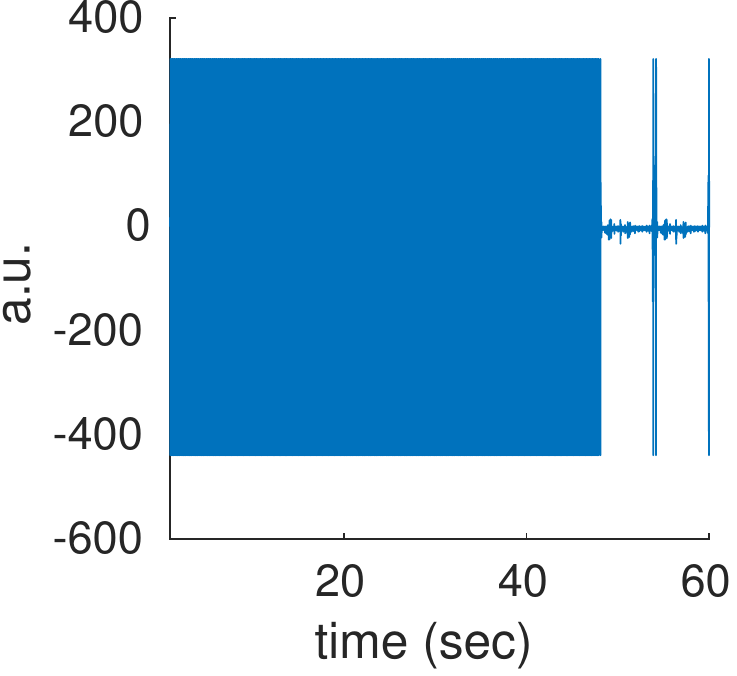}\label{fig:high-noise}}
     \hfill
     \caption{Examples of tested noise signals.}
     \label{fig:noise-examples}
\end{figure}

\subsection{Phonemes Detection Accuracy}
\label{sec:phon_cross}

We then evaluate the accuracy of detecting each phoneme for each speaker
as well as the false positive across phonemes and speakers. 
In particular, we run \name to match each \acc sample
(88 samples --- corresponding to each phoneme and speaker)
to all the collected \mic samples; each \acc sample must match
only one \mic sample. Table~\ref{table:phonetic_eval} shows the matching results.

First, we match the consonant phonemes across the two speakers as evident
in the first row. The true positive rate exceeds 90\%, showing that \name can
correctly match the vast majority of consonant
phonemes. This is analogous to a low false negative rate of less
than 10\%. Moreover, we report the false positive rate
which indicates the instances where \name matches an \acc sample 
to the inappropriate \mic sample. As shown in
the same figure, the false positive rate is nearly zero.

Having such a very low false positive rate highlights two 
security guarantees that \name offers. It does not mistake
a phoneme as another even for the same speaker. Recall that 
pitch information is widely used to perform
speaker recognition, as each person has
unique pitch characteristics that are independent
of the speech. \name overcomes pitch characteristics
similarity and is able to distinguish the different phonemes
as spoken by the same speaker. 

Moreover, \name successfully distinguishes the same phoneme
across the two speakers. A phoneme contains speaker-independent features. 
\name overcomes these similar features to effectively identify each 
of them for each speaker. The fourth row of 
Table~\ref{table:phonetic_eval} shows comparable results when 
attempting to match the vowel phonemes for both speakers.
 
The second and third rows complete the picture of phoneme matching. 
They show the matching results of the vowel phonemes to the consonant 
phonemes for both speakers. Both
rows do not contain true positive values as there are no phoneme
matches. Nevertheless, one must notice the very low false positive
ratio that confirms the earlier observations. Finally, the fifth row 
shows results of matching all the \acc samples to all the \mic samples. The 
true positive rate is 93\%, meaning that \name correctly matched 82 \acc samples 
matched to their \mic counterparts. Moreover, the false positive rate was only 0.6\%.

\subsection{Idle Detection Accuracy}
\label{sec:fp}

Last but not least, we evaluate another notion of false positives:
\name mistakenly matches external speech to a silent user.
We record idle (the user not actively
speaking) segments from \name's \acc and attempt to match them to the 
recorded phonemes of both participants. We considered two types of 
idle segments: the first contains no energy from speech or other 
movements (Fig.~\ref{fig:low-noise}), while the other contains significant abrupt
motion of the \acc resulting in recordings with high energy spikes
(similar to the spike of Fig.~\ref{sig-raw}). We also constructed a high energy noise 
signal with periodic patterns as shown in Fig.~\ref{fig:high-noise}.

We execute \name over the different idle segments and
\mic samples and recorded the false matching decisions. 
In all of the experiments, we did not observe any occurrence of a 
false matching of an idle \acc signal to any phoneme
from the \mic for both speakers. As recorded
phonemes are representative of all possible sounds 
comprising the English language, we can be confident
that the false positive rate of \name is zero in practice
for silent users.

\section{Theoretical Analysis}
\label{sec:model}

In this section, we highlight the effectiveness of the per-segment analysis of \name's matching 
algorithm in preventing an attacker from injecting commands. We also show 
that our matching algorithm ensures the phonetic-level results 
constitute a lower-bound of sentence-level matching. We provide a formal analysis of 
this property, which will be further supported in the next section using real user studies.

\subsection{Model}

We analyze the properties of \name's matching algorithm which 
takes as inputs two signals $f(t)$ and $g(t)$ originating from
the \acc and \mic, respectively. It outputs a final matching result 
that is a function of normalized cross-correlation of $f(t)$ and $g(t)$:
$h(t) = \frac{f(t) \star g(t)}{E}$, where $E = \sqrt{\lVert f(t) \rVert.\lVert g(t)\rVert}$,
$\star$ denotes the cross-correlation operator, and $\rVert \cdotp \lVert$
is the energy of the signal (autocorrelation evaluated at 0).
For the simplicity of the analysis, we will focus on the most important feature of $h(t)$, 
its maximum value. We can then define \name's binary matching function, $v(f,g)$, as:
\begin{equation}
\label{eq:rule}
v(f,g) = 
     \begin{cases}
      {v=0,} &\quad\text{if \,} 0 \le  m=\max(|h(t)|) \le th \\
       {v=1,} &\quad\text{if \,} th < m=\max(|h(t)|) \le 1. \
     \end{cases}
\end{equation} 

Each of the input signals comprises a set of segments, which could refer to the phonemes 
making up a word or words making up a sentence, depending on the granularity of the analysis. 
Let  $f_i(t)$ and $g_i(t)$ be the $i^{th}$ segments of $f(t)$ and $g(t)$, respectively. 
We assume that maximum length of a segment is $\tau$,
such that $f_i(t)=0, t \in (-\infty , 0] \bigcup [\tau, +\infty)$.
We can then rewrite $f(t)$ as $f(t) = \sum_{i=1}^{n}{f_i(t-i\tau})$;  
the same applies for $g(t)$. 

One can view the cross-correlation operation as sliding the segments $g_i(t)$ 
of $g(t)$ against those of $f(t)$. 
The cross correlation of $g(t)$ and $f(t)$ can be computed as:
\begin{dmath*}
{h_c(t) = f(t) \star g(t) = }\\
\sum_{i=1}^{n-1}\left(\sum_{j=1}^{i}(f_{j}\star g_{n-i+j} (t-(i-1)\tau))\right)\\
+\sum_{k=1}^{n}(f_{k}\star g_{k} (t-(n-1)\tau))\\
+\sum_{i=n-1}^{1}\left(\sum_{j=1}^{i}(f_{n-i+j}\star g_{j} (t-(2n-i-1)\tau))\right).
\end{dmath*}
The normalized cross correlation, $h(t)$, is obtained by normalizing $h_c(t)$ to 
$E=\sqrt{\lVert\sum_{i=1}^{n}{f_i(t-i\tau)}\rVert.\lVert\sum_{i=1}^{n}{g_i(t-i\tau})\rVert}$. 
Since the segments of $f$ and $g$ do not overlap each other (by their definition), 
the energy of a signal is the sum of the energies of its components, such that 
$E=\sqrt{\sum_{i=1}^{n}{\lVert f_i(t)\rVert}.\sum_{i=1}^{n}{\lVert g_i(t}\rVert)}$. 
Finally, we expand $E$ to obtain the final value of the normalized cross correlation between $f$ and $g$ as:
\begin{equation}
h(t) = \frac{h_c(t)}{\sqrt{\sum_{i=1}^{n}\sum_{j=1}^{n}{\lVert f_i(t)\rVert . \lVert g_j(t)\rVert}}}.
\end{equation}

To decide on the final outcome, \name computes $\max{\abs {h(t)}}$, 
which, according to the triangle rule, becomes:
\begin{equation}
\label{eq:cross}
\max{\abs{h(t)}} \le 
\max {\frac{|h_c(t)|}{\sqrt{\sum_{i=1}^{n}\sum_{j=1}^{n}{\lVert f_i(t)\rVert . \lVert g_j(t)\rVert}}}}.
\end{equation}

We assume that the segments' cross-correlation maximizes when they are aligned. 
That is, $max(f_i \star g_j) = \sum\limits_{t=0}^\tau f_i(t)g_i(t)$; otherwise, we can 
redefine the segments to adhere to this property. We can then separate the components
of Eq.~\eqref{eq:cross} into different components such as:
\begin{dmath}
\label{eq:max}
\max{\abs{h(t)}} \le \frac{1}{E}. \max (h_l, h_m,h_r), \text{where} \\ 
{h_l = \max_{i=1\ldots n-1}\left(\sum_{j=1}^{i}|f_{j}\star g_{n-i+j} (t)|\right),}\\
{h_m = \max{|f_{k}\star g_{k}(t)|}, \text{and}}\\
{h_r = \max_{i=1\ldots n-1}\left(\sum_{j=1}^{i}|f_{n-i+j}\star g_{j} (t)|\right).}
\end{dmath}

The above equation describes how the final outcome is related to the results of 
running \name on the segments comprising $f(t)$ and $g(t)$. 
Two segments $f_{i}(t)$ and $g_{j}(t)$ are positively matched when their 
maximum of normalized cross correlation, $m_{ij}$, is between $th$ and 1. 
Otherwise, there is a negative match.

The value of $m_{ij}$ can be given as:
\begin{equation}
\label{eq:ind}
m_{i,j} = \max{\frac{\abs{f_{i}\star g_{j} (t)}}{\lVert f_{i}(t)\rVert . \lVert g_{j} (t)\rVert}}. 
\end{equation}

Let $e_{i,j}$ denote the product of the energies of $f_{i}$ and $g_{j}$,
such that $e_{i,j} = \lVert f_{i}(t)\rVert . \lVert g_{j} (t)\rVert$. 
After applying the triangle rule to Eq.~\eqref{eq:max}, the final 
outcome $m = \max{\abs{h(t)}}$ can be given as:
\begin{dmath}
\label{eq:analyze}
m \le \frac{1}{E}.\max 
\left(
{\max\limits_{i=1\ldots n-1}{\sum\limits_{j=1}^{i} {m_{i,n-i+j}.e_{i,n-i+j}}},}\\
{\sum\limits_{k=1}^{n} {m_{kk}.e_{kk}}},
{\max\limits_{i=1\ldots n-1}{\sum\limits_{j=1}^{i} {m_{n-i+j,j}.e_{n-i+j,j}}}}.
\right)
\end{dmath}

It is evident from Eq.~\eqref{eq:analyze} how the final outcome of \name depends
on computing the maximum of $2n-1$ distinct components. 
Each component, $\sum\limits_{j=1}^{i} {m_{ij}.e_{ij}}$, is simply the weighted average of 
the outcomes of \name when it matches the included segments.
Without loss of generality, let's consider the case when $n=2$:
\begin{equation}
\label{eq:analyzetwo}
m \le 
\max{\left(
\frac{m_{11}.e_{11}}{E}, 
\frac{m_{12}.e_{12}+m_{21}.e_{21}}{E},
\frac{m_{22}.e_{22}}{E}
\right)},
\end{equation}
where $m_{ij}$ are as defined in Eq.~\eqref{eq:ind}; $m_{11}$ and $e_{11}$ are the results
of matching $f_1$ and $g_2$, $m_{12}$ and $e_{12}$ are those of $f_1$ and $g_1$,
$m_{21}$ and $e_{21}$ are those of $f_2$ and $g_2$, and 
$m_{22}$ and $e_{22}$ are those of $f_2$ and $g_1$.

Given the above model of the final outcome of \name as a function
of the segments composing the commands, we study the properties  of
\name as described below.

\subsection{Per-segment Analysis}

Eq.~\eqref{eq:analyzetwo} reveals the importance of the per-segment
analysis of \name. This step thwarts an attacker's ability to 
inject commands into the voice assistant system. The attacker aims
to inject segments to $g(t)$ that do not bring $m$ below $th$ (so that
\name generates a positive match according to Eq.~\eqref{eq:rule}).
If there were no per-segment
analysis, an attacker could exploit matching segments to inject segments
that do not match. The middle component of Eq.~\eqref{eq:analyzetwo} explains it. 
Assuming that $m_{1,2}.e_{1,2}$ is large enough, the attacker
can inject $g_2(t)$ such that $m_{1,2}.e_{1,2} + m_{2,1}.e_{2,1}$ is still
large, despite  $m_{2,1}$ being low. This happens when $e_{2,1}$ is too
low, implying that the \acc did not record the injected segment.

The per-segment analysis of \name addresses this issue using three mechanisms.
\textit{First}, it removes all portions of $f(t)$ that fall below the running average of
the noise level. These removed portions will not even be part of 
Eq.~\eqref{eq:analyzetwo}. So, the attacker cannot inject commands when
the user is silent (no corresponding \acc signal).  \textit{Second}, if the energy of some segment 
of $f(t)$ is above the noise level, \name normalizes its magnitude to 1, 
after removing the spikes. As such, it aims to make the energies
of the segments of $f(t)$ uniform. The attacker cannot inject a command
with very low energy as it will not be recorded by the \mic of the voice
assistant. This forces $e_{2,1}$ to be comparable to $e_{1,2}$. 
As a result, a low value of $m_{2,1}$ reduces the value of
$m$ of Eq.~\eqref{eq:analyzetwo}.\textit{Third}, and more importantly,
The per-segment analysis of \name nullifies those segments which 
have their maximum normalized cross-correlation falling below a threshold (equal to 0.4
in our evaluation). These segments will not make it to the final decision stage, and will not 
be part of Eq.~\eqref{eq:analyzetwo}.

\subsection{False Positive Rate}

The results of Section~\ref{sec:phonetics} show that the false positive 
rate of matching is not zero for the English phonemes. Such a false positive
rate opens up security holes in \name. We show below that while the false
positive rate is not zero at the phonetic level, adding more phonemes
to the command will drive the false positive rate closer to zero.
In other words, the more sounds the user speaks (i.e., the longer the command is),
the lower the false positive rate will be. 

To show this, we will take another look at Eq.~\eqref{eq:analyze},
where $f_i$ and $g_i$ represent the phonemes making up the command.
At the phonetic level, a false positive event occurs when $m_{i,j}>th$, given that
$f_{i}$ does not match $g_{j}$. 
As evident from Eq.~\eqref{eq:analyze}, when the values of $e_{i,j}$
are roughly uniform which we ensure from the per-segment analysis,
the value of $m$ is simply an average of the values of $m_{i,j}$.
The final matching result $v(f,g)$, is a direct function of $m$. 
A false positive event occurs when $v(f,g)=1$ or $m>th$, given that the underlying
\acc ($f(t)$) and \mic ($g(t)$) signals do not match. 

There are two cases available in Eq.~\eqref{eq:analyze}: $i <n$ 
(the first and third terms in the max) and $i=n$ (the middle term in the max).
The former case is simple; the final value of $m$
is by definition a scaled-down version of the $m_{ij}$s. 
A lower value of $m$ will lower the false positive rate.
  
The latter case considers $n$ segments (phonemes) composing the command.
A false positive event occurs when $m=1/n\sum\limits_{k=1}^{n}{m_{k,k}}>th$, 
given that $f(t)$ does not match $g(t)$.
In the case of phonemes false positives, one can view all $m_{i,j}$s as being drawn from the distribution 
$P_M(m_{i,j}) = P_M(M=m_{i,j}|f_{i} \neq g_{j})$, where the $\neq$ operator indicates non-matching; 
the false positive rate is simply $P_M(M>th|f_{i} \neq g_{j})$.
The false positive rate of the whole command is then equal to
$P_M(m) = P_M(\sum_{k=1}^n{m_{k,k}/n}>th|f \neq g)$.
Our objective reduces to showing that $P_M(m)$ decays as $n$ increases
(i.e., more non-matching phonemes are added to the command). 

The distribution $P_M(M=m_{ij})$ is an arbitrary one that only 
satisfies two conditions. First, it is bounded since the values of
$m_{i,j}$ are limited to the interval $[0,1]$. Second, the matching
threshold, $th$, is larger than the mean of the distribution
such that $th>E(m_{i,j})$. The empirical false positive distribution  $P(m_{i,j})$ 
that we estimated in Section~\ref{sec:phonetics} satisfies both conditions. 

We know from the Hoeffding bound that since $m_{i,j}$ are bounded,
$P_M(\sum_{k=1}^n{m_{k,k}}-n.E(m)>t)\le e^{\frac{-2t^2}{n}}$, for $t \ge 0$. 
Substituting $t = n.th-n.E(m)$ (which is larger than 0) yields: 
\begin{equation}
\label{eq:decay}
P_M(\frac{1}{n}.\sum_{k=1}^n{m_{nk}}>th)\le e^{-2n(th-E(m))^2}.
\end{equation}

The left-hand side of Eq.~\eqref{eq:decay} is simply the false 
positive rate of the command composed of $n$ non-matching phonemes.
Clearly, this false positive rate decays exponentially fast in $n$. 
Our results from the user study further confirm this analysis. 

\section{Evaluation}
\label{sec:evaluation}

We now evaluate the efficacy of \name in identifying common voice assistant commands, 
under different scenarios and for different speakers. We demonstrate 
that \name delivers almost perfect matching accuracy (True Positives, TPs) regardless 
of its position on the body, user accents, mobility patterns, or even across different languages. 
Moreover, we elaborate on the security properties of \name, demonstrating 
its effectiveness in thwarting various attacks. Finally, we report the 
delay and energy consumption of our wearable prototypes.

\subsection{User Study}

To support the conclusions derived from our model, we conducted a detailed user study 
of the \name prototype with 18 users and under 6 different scenarios. We 
tested how \name performs at three positions, each corresponding to a different 
form of wearable (Fig.~\ref{fig:scenarios}) eyeglasses, earbuds, and necklace. 
At each position, we tested two cases, 
asking the user to either stand still or jog. In each scenario, We asked the participants 
to speak 30 phrases/commands (listed in Table~\ref{table:commands} of Appendix A). 
These phrases represent common commands issued to the ``Google Now" voice assistant. 
In what follows, we report \name's detection accuracy (TPs) and false 
positives (FPs) when doing a pairwise matching of the commands for each participant.
We collected no personally identifiable information from the individuals, 
and the data collection was limited to our set of commands and posed
no privacy risk to the participants. As such, our user study meets the IRB 
exemption requirements of our institution.

\paragraph*{\textbf{Still}}

\begin{figure}[t]
     \centering
     \subfigure[earbuds]{\includegraphics[width=0.32\columnwidth]{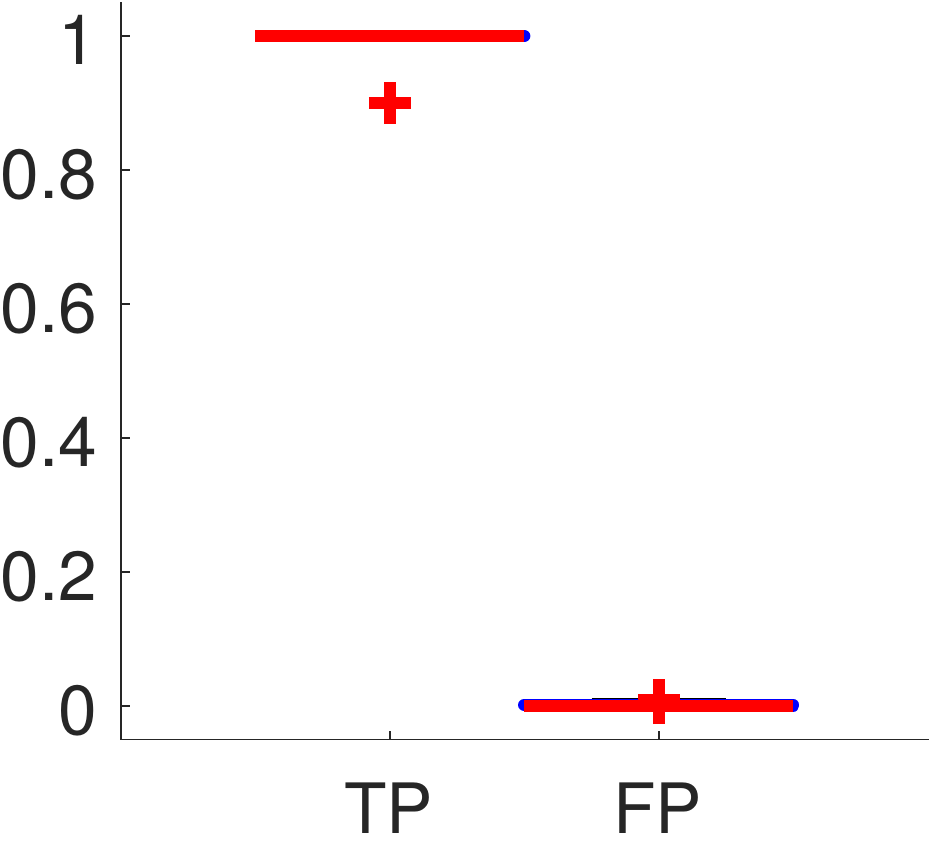}\label{apple-still}}
     \hfill
     \subfigure[eyeglasses]{\includegraphics[width=0.32\columnwidth]{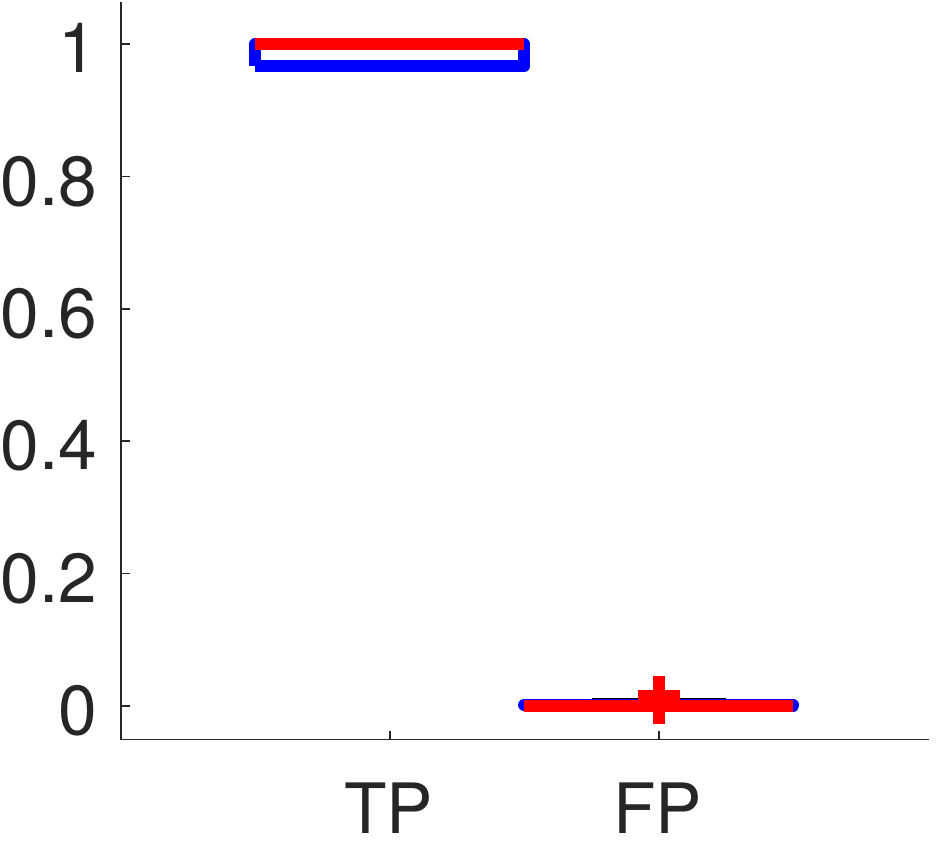}\label{google-still}}
     \hfill  
     \subfigure[necklace]{\includegraphics[width=0.32\columnwidth]{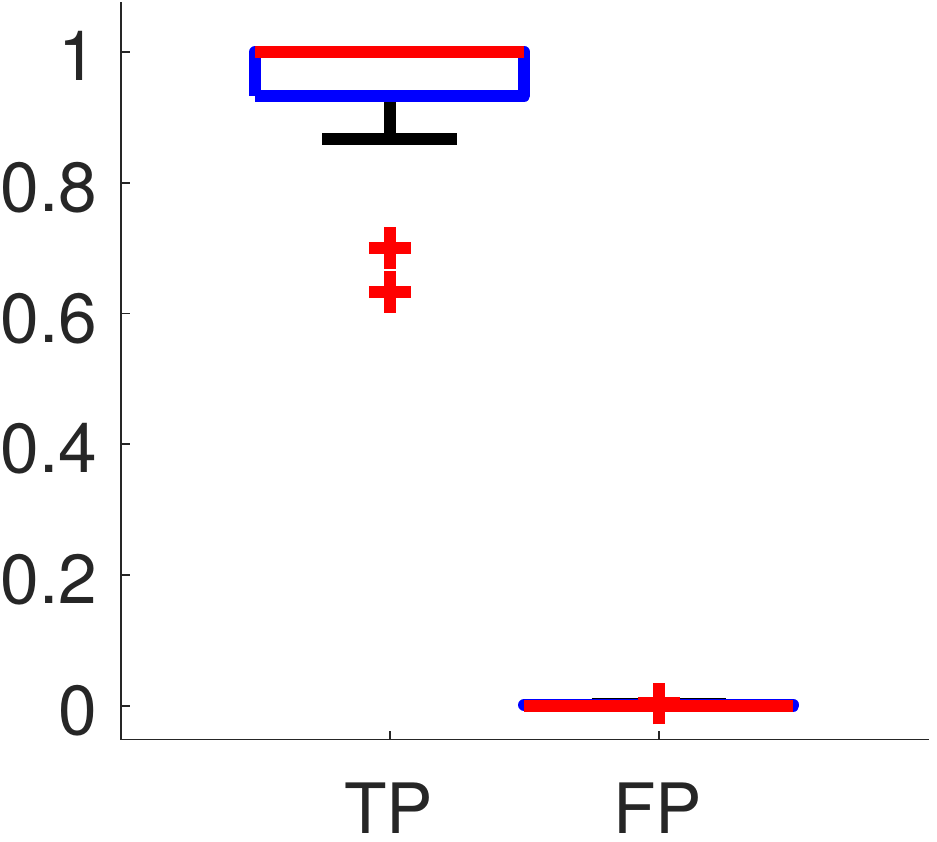}\label{neck-still}}
     \caption{The detection accuracy of \name for the 18 users in the still position.}
     \label{fig:match-still}
\end{figure}

\begin{figure}[t]
     \centering
     \subfigure[User A]{\includegraphics[width=0.47\columnwidth]{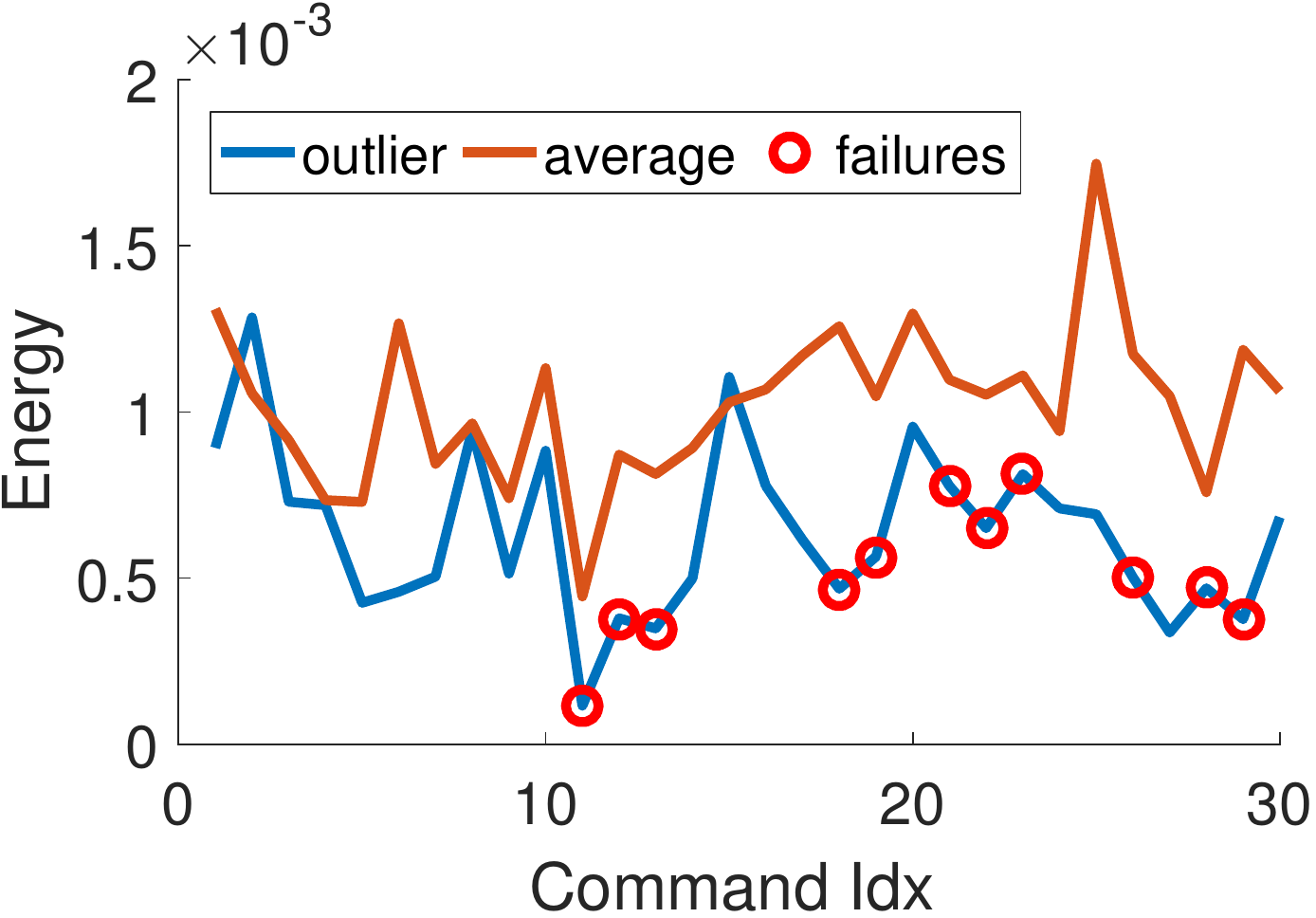}\label{outlier_mic}}
     \hfill  
     \subfigure[User B]{\includegraphics[width=0.47\columnwidth]{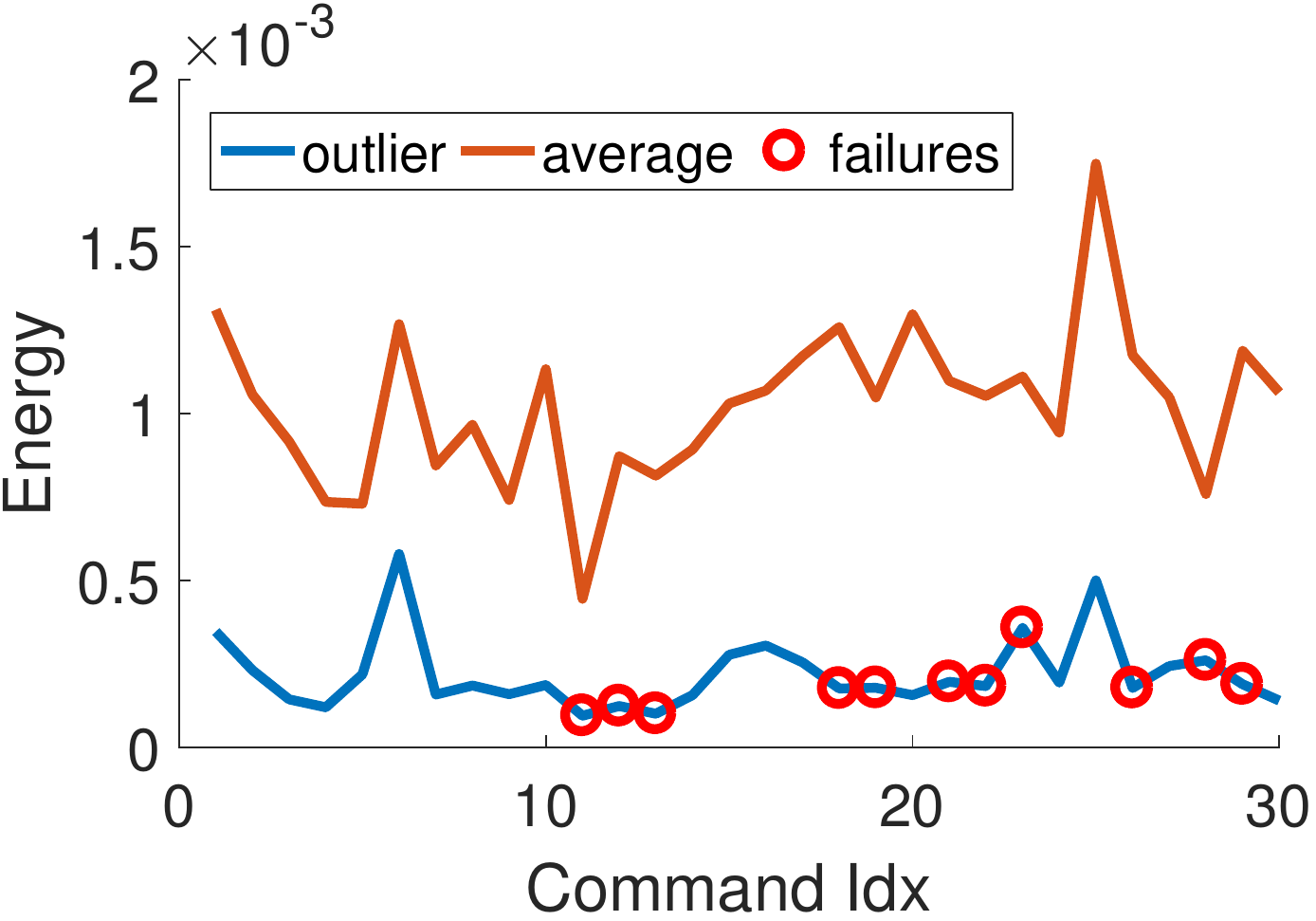}\label{outlier_mic_2}}
     \caption{The energy levels of the outlier users (in Fig.~\ref{neck-still}) compared to 
     average users. The circles represent commands of the outlier users that \name fails to match.}
     \label{fig:match-failure}
\end{figure}

\name delivers high detection accuracy (TPs), with the overall accuracy 
rate very close to 100\% (more than 97\% on average). This indicates most of the commands are 
correctly authenticated from the first trial and \name does not introduce a usability burden to the user. 
The false positive rate is 0.09\% on average, suggesting that very few signals will leak 
through our authentication. These false positive events occur because the per-segment analysis 
of our matching algorithm removes all non-matching segments from both signals, which ensures the 
security properties of \name. In these cases, when the remaining segments for the \mic signal
accidentally match what the user said and leak through \name, the voice recognition 
system (Voice-to-Text) fails to pick them up as sensible voice commands. 
Fig.~\ref{fig:match-still} shows the overall distribution of detection results for each scenario.

\name performs almost perfect in two wearable scenarios, eyeglasses and earbuds, but has 
two outliers regarding the detection accuracy in the case of the necklace. We looked into the 
commands that \name fails to recognize and found they happen when there are significant 
energy dips in the voice level. Fig.~\ref{fig:match-failure} reports the energy levels of 
the voice sessions for our two outlier users compared to the average across users.
This suggests both participants used a lower (than average) voice when doing the experiments
which did not generate enough energy to ensure the authentication.

\begin{figure}[t]
     \centering
     \subfigure[earbuds]{\includegraphics[width=0.32\columnwidth]{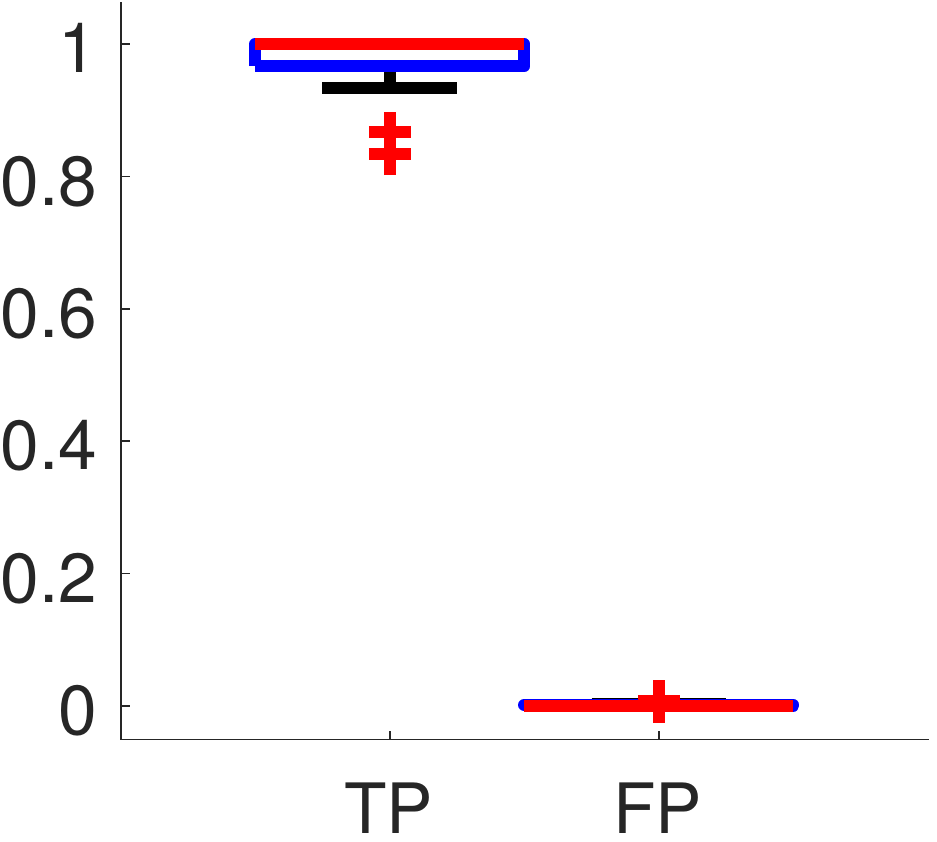}\label{apple-jump}}
     \hfill
     \subfigure[eyeglasses]{\includegraphics[width=0.32\columnwidth]{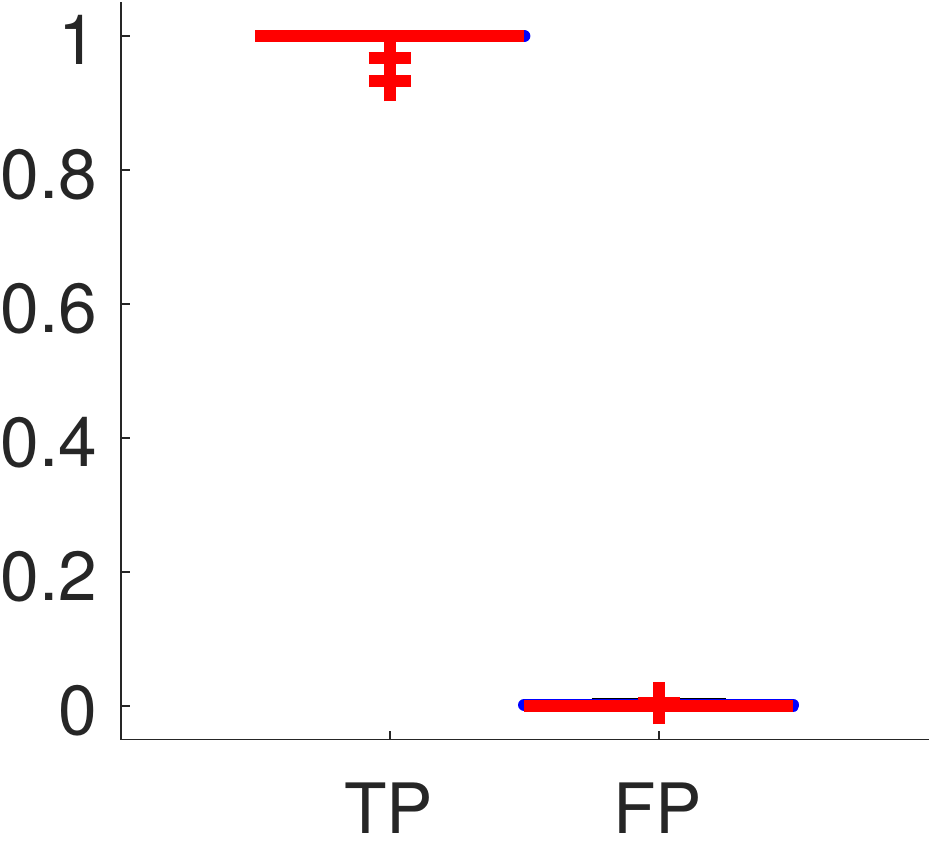}\label{google-jump}}
     \hfill       
     \subfigure[necklace]{\includegraphics[width=0.32\columnwidth]{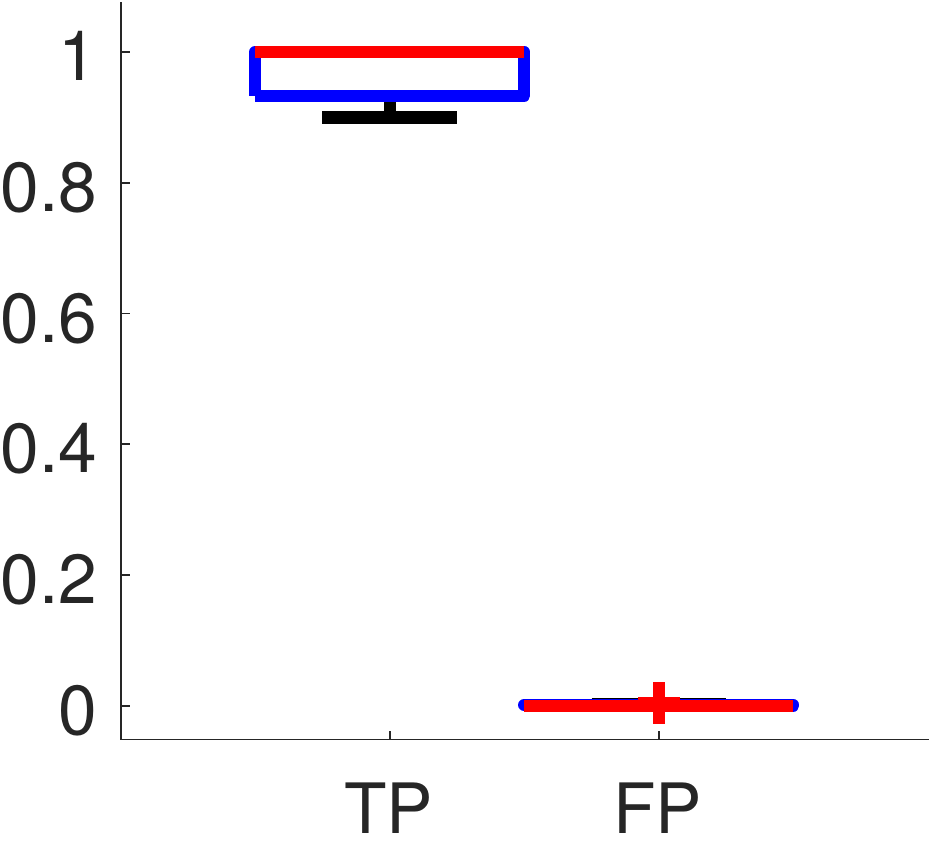}\label{neck-jump}}
     \caption{The detection accuracy of \name for the 18 users in the moving position.}
     \label{fig:match-jump}
\end{figure}

\paragraph*{\textbf{Mobility}}

We asked the participants to repeat the experiments at each position while jogging. 
Our algorithm successfully filters the disturbances introduced by moving, breathing 
and \name's match accuracy remains unaffected (see Fig.~\ref{fig:match-jump}). In fact, 
we noticed in certain cases, 
such as for the two outliers observed in our previous experiments, the results are even 
better. We studied the difference between their samples in the two scenarios and found 
both \acc and \mic received significantly higher energy in the jogging scenario even 
after we filtered out the signals introduced by movement. One explanation is users 
are aware of the disturbance introduced by jogging and try to use louder voice to 
compensate. This observation is consistent across most of our participants, not 
just limited to the two outliers.

\begin{figure*}[t]
     \centering
     \subfigure[earbuds]{\includegraphics[width=0.6\columnwidth]{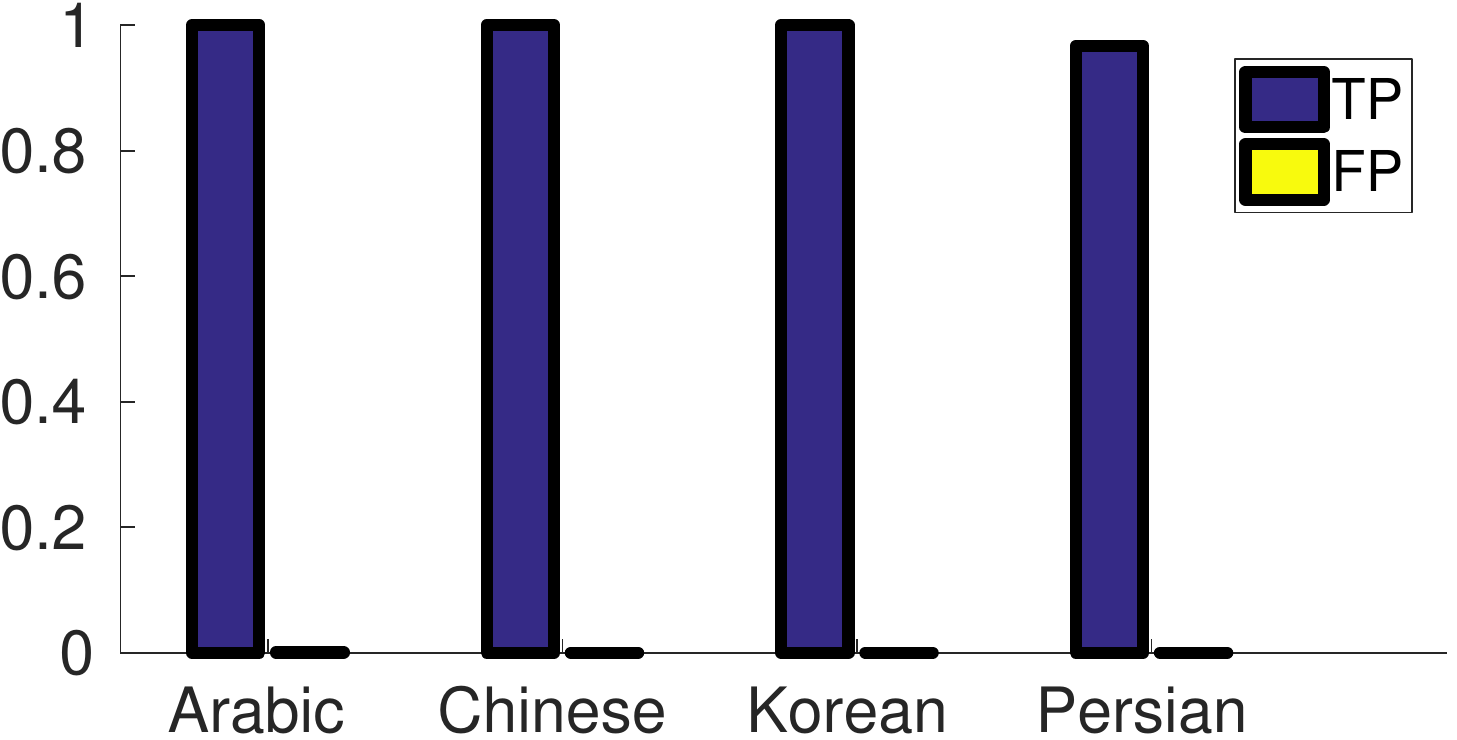}\label{languages-apple-still}}
     \hfill
     \subfigure[eyeglasses]{\includegraphics[width=0.6\columnwidth]{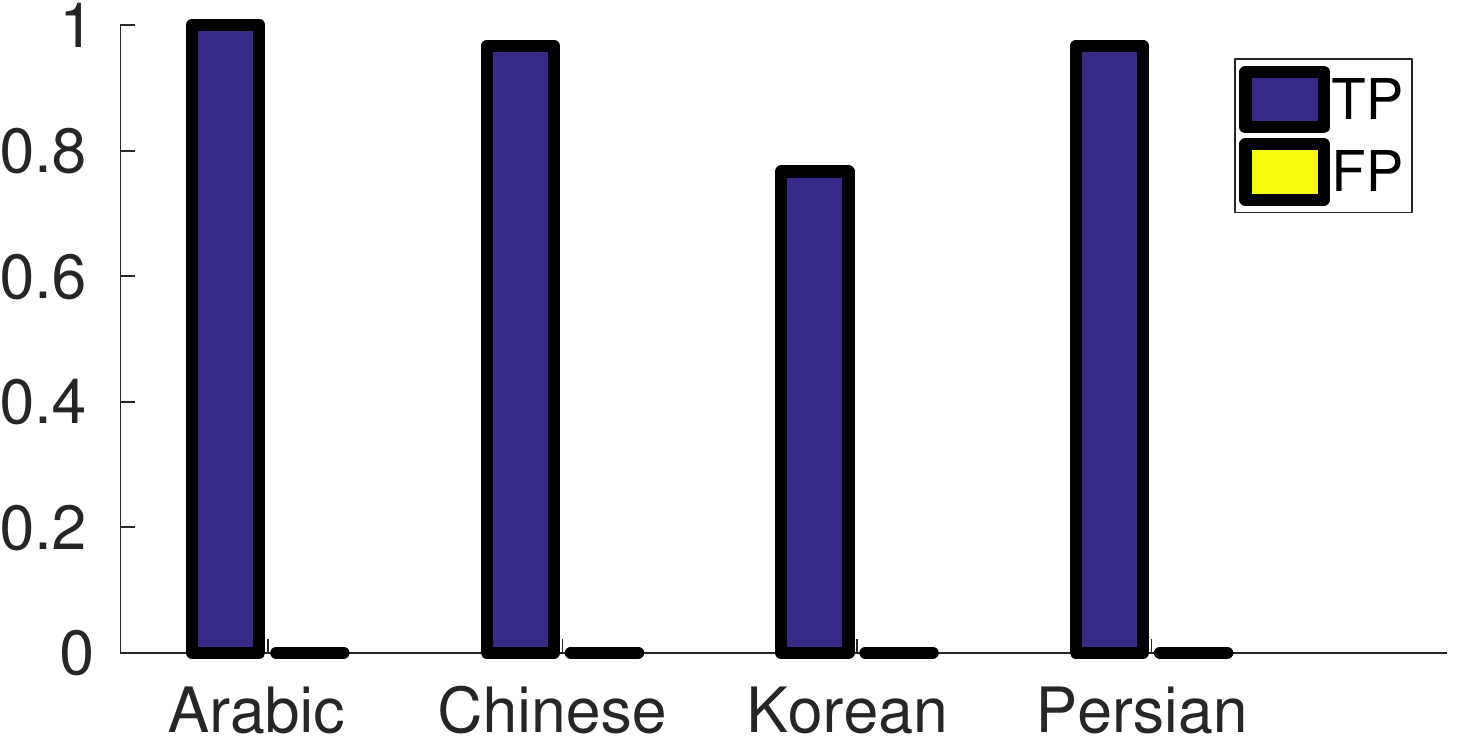}\label{languages-google-still}}
     \hfill
     \subfigure[necklace]{\includegraphics[width=0.6\columnwidth]{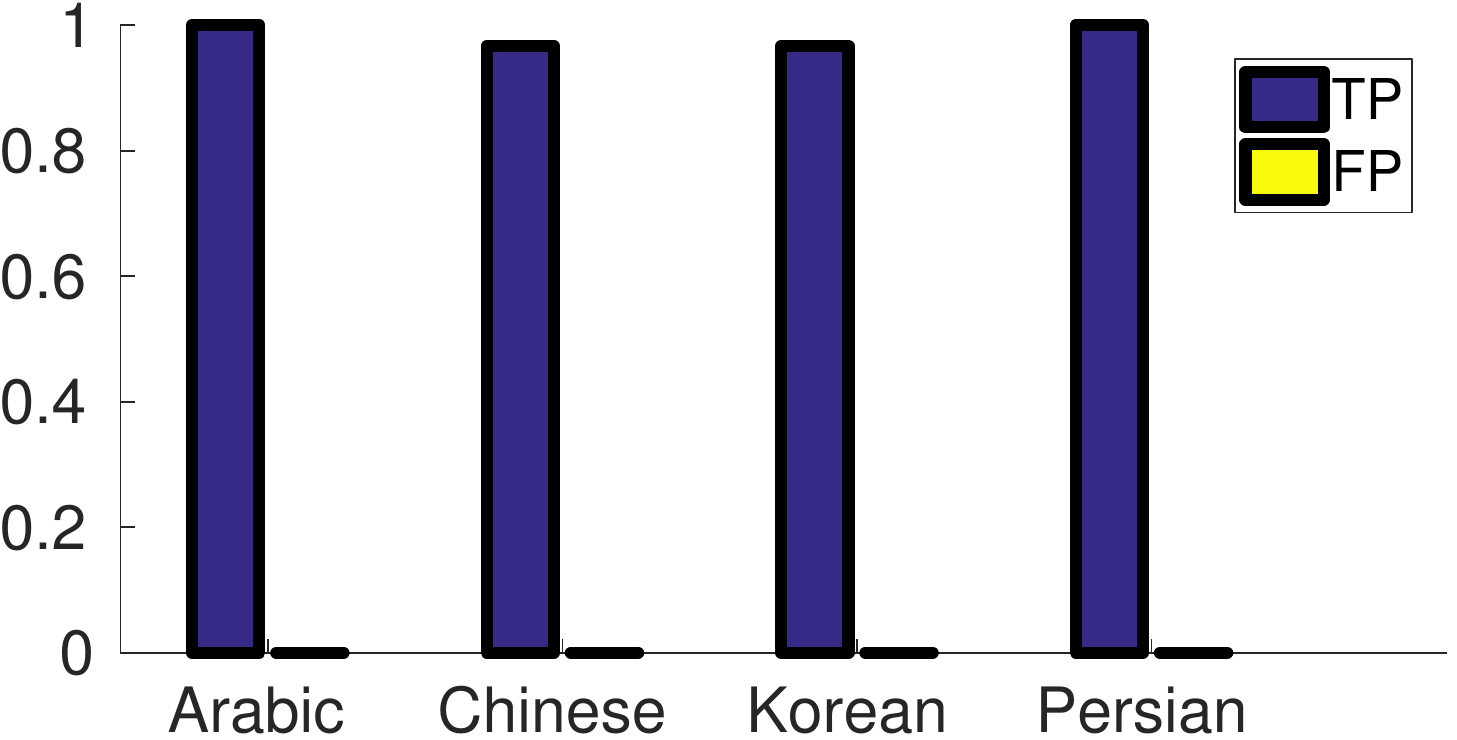}\label{languages-neck-still}}
     \caption{The detection accuracy of \name for the 4 different languages.}
     \label{fig:languages-still}
\end{figure*}

\paragraph*{\textbf{Language}}
We translated the list of 30 commands into four other languages --- Arabic, Chinese, Korean 
and Persian --- and recruited four native speakers of these languages. We asked the participants 
to place and use \name at the same three positions. As shown in Fig.~\ref{fig:languages-still}, 
\name performs surprisingly well, even though the \name prototype was trained on English phonemes
(Section~\ref{sec:matching_decision}). 
\name delivers almost perfect detection accuracy, except for one case, with the user 
speaking Korean when wearing eyeglasses. The Korean language lacks nasal consonants, and 
thus does not generate enough vibrations through the nasal bone~\cite{yoshida2008phonetic}.

\subsection{Security Properties}

In Section~\ref{sec:system-threat-models}, we listed three types of
adversaries against which we aim to protect the voice assistant systems. 
\name can successfully thwart attacks by these adversaries through 
its multi-stage matching algorithm. Table~\ref{table:features} lists 
the protections offered by \name when the user is silent and actively speaking. 
Here, we use the evaluation results in Section \ref{sec:evaluation} to elaborate 
on \name's security features for each attack scenario and both cases when 
the user is silent and speaking.

\begin{table}[t]

\renewcommand{\arraystretch}{1.1}
\footnotesize
  \centering
   \caption{The protections offered by \name.}
\begin{tabular}{L{0.5cm} C{1.7cm} C{2.2cm} C{1.2cm}  C{1.2cm}}
\toprule
\textbf{Scenario} & \textbf{Adversary} & \textbf{Example} & \textbf{Silent User} 
& \textbf{Speaking User}\\ 

\midrule 

A & Stealthy & mangled voice, wireless-based & \tick & \tick \\
B & Biometric Override & replay, user impersonation & \tick & \tick \\
C & Acoustic Injection & direct communication, loud voice & distance cut-off & distance cut-off \\

\bottomrule
\end{tabular} 

\label{table:features}
\end{table}

\paragraph*{\textbf{Silent User}}
When the user is silent, \name completely prevents 
any unauthorized access to the voice assistant. In Section~\ref{sec:fp}, 
we evaluate the false positive rate of \name mistakenly classifying noise while 
the user is silent for all English phonemes. We show that \name has a zero false positive
rate. When the user is silent, the adversary \textit{cannot inject} 
any command for the voice assistant, especially for scenarios A and B 
of Section~\ref{sec:system-threat-models}. There is an exception, 
however, for scenario C; an adversary can employ a very loud sound to 
induce vibrations at the \acc chip of \name. Note that, since the \acc 
only senses vibrations at the z-axis, the attacker must make the extra 
effort to direct the sound wave perpendicular to the \acc sensing surface.
Next, we will show that beyond a cut-off distance of 30cm, very loud sounds 
(directed at the z-axis of the \acc)
do not induce \acc vibrations. Therefore, to attack \name, an adversary has 
to play a very loud sound within less than an arm's length from the 
user's body --- which is highly improbable. 

\begin{figure}[t]
     \centering
     \subfigure[Exposed \acc]{\includegraphics[width=0.45\columnwidth]{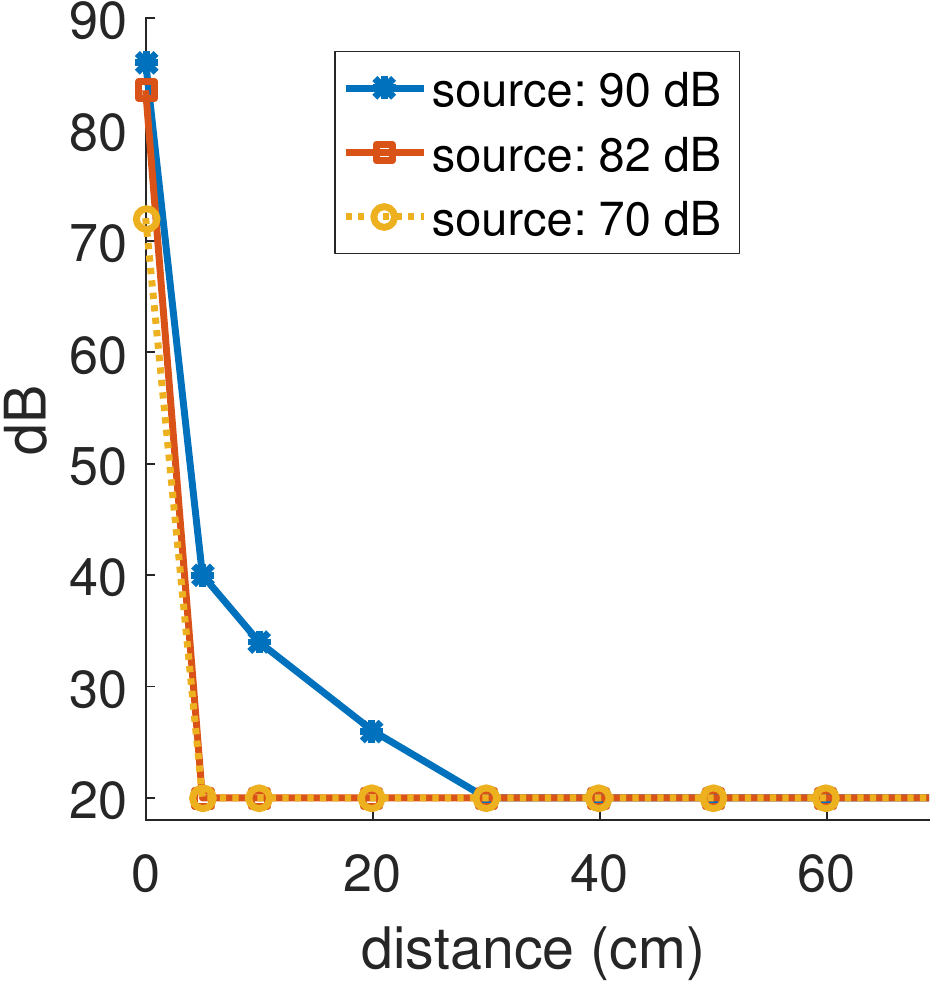}
     \label{fig:exposed-distance}}
     \hfill  
     \subfigure[Covered \acc]{\includegraphics[width=0.45\columnwidth]{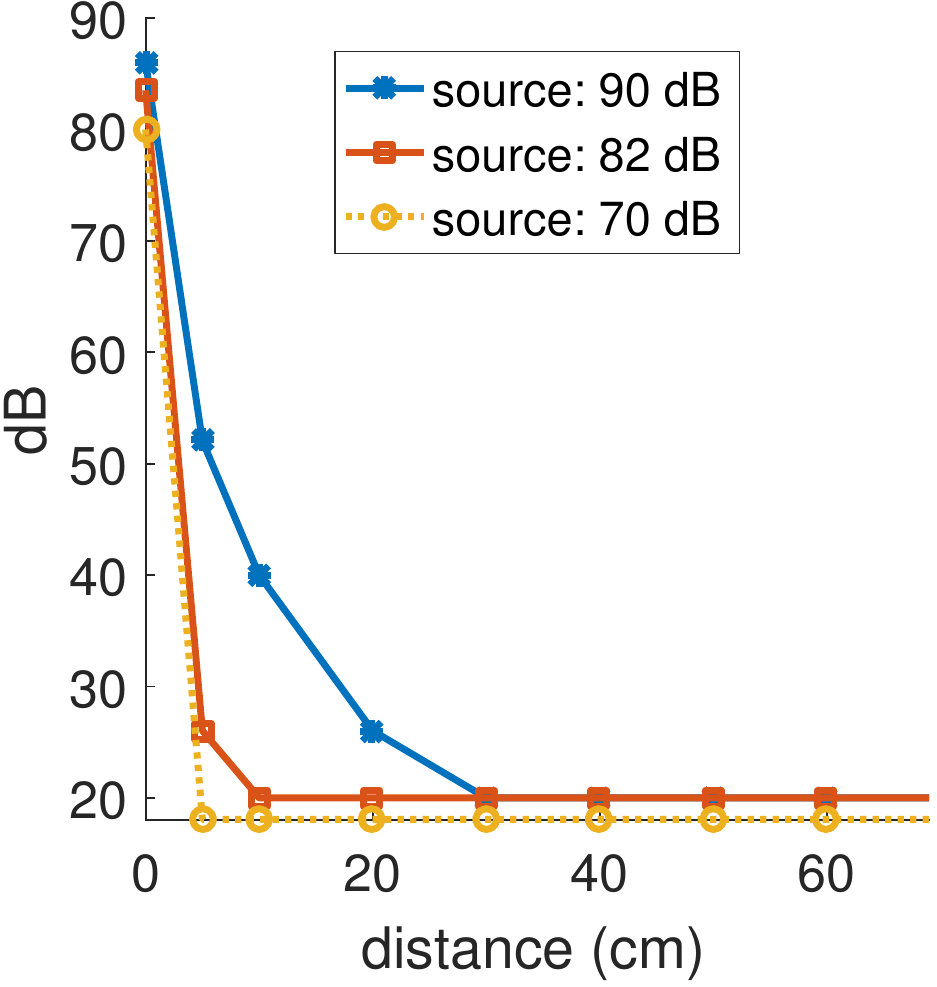}
     \label{fig:covered-distance}}
     \hfill
     \caption{The magnitude of the sensed over-the-air vibrations by the \acc as a function of the distance between the sound source and the \acc.}
     \label{fig:acc-distance}
\end{figure}

We conduct experiments on the cut-off distances in two scenarios: the first 
with \name exposed and the second with \name covered with cotton clothing. Fig.~\ref{fig:acc-distance} 
reports how the \acc chip of \name reacts to over-the-air sound signals 
of different magnitudes at different distances. In each of these scenarios, 
we played a white noise at three sound 
levels:\footnote{\url{http://www.industrialnoisecontrol.com/comparative-noise-examples.htm}}
2x, 4x and 8x the conversation level at 70dB, 82db and 90dB, respectively. 
The noise is directed perpendicularly to the sensing surface of the \acc.
Fig.~\ref{fig:exposed-distance} shows the recorded magnitude
of the \acc signal as a function of the distance between 
the sound source and \name when it is exposed.
As evident from the plots, there is a cut-off distance of 30cm, where
\name's \acc cannot sense even the loudest of the three sound sources.
For the other two sounds, the cut-off distance is 5cm. Beyond the 
cut-off distance, the magnitude of the recorded signal is the same as that
in a silent scenario. This indicates that an adversary cannot 
inject commands with a high sound level beyond some cut-off distance. 
These results are consistent with the case of \name covered with cotton, 
as shown in Fig.~\ref{fig:covered-distance}. The cut-off
is still 30cm for the loudest sound. It is worth noting that the recorded
signal at the \mic does not change magnitude as drastically as a function 
of the distance. At a distance of 1m from the sound source, the audio signal 
loses at most 15dB of magnitude.

\paragraph*{\textbf{Speaking User}}
On the other hand, the adversary may try to launch an attack
on the voice assistant system when the user is actively speaking.
Next, we show how \name can successfully thwart the stealthy attacks 
in scenario A. We will show, in the most extreme case scenario, how 
\name can completely distinguish the \acc samples of the voice spoken 
by the user from the reconstructed sound of the same command, even when the 
reconstructed voice sounds the same to the human listener as the original one.

Vaidya \textit{et al.}~\cite{vaidya2015cocaine,carlini2016hidden} presented an attack 
that exploits the gap between voice recognition system and human voice 
perception. It constructs mangled voice segments that match the MFCC 
features of an injected voice command. An ASR engine can recognize the 
command, but not the human listener. This and similar attacks rely on 
performing a search in the MFCC algorithm parameter space to find 
voice commands that satisfy the above feature.

\begin{figure}
     \centering
     \includegraphics[width=0.9\columnwidth]{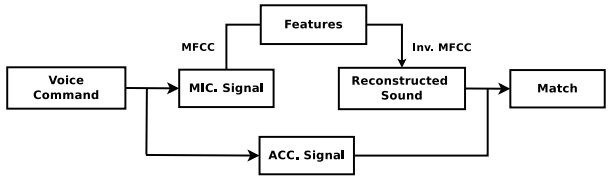}
     \caption{The flow of the mangling voice analysis.}
     \label{fig:mfcc-exp}
\end{figure}

Performing an exhaustive search on the entire parameter space of the 
MFCC generation algorithm is prohibitive. Therefore, to evaluate 
the effectiveness of \name against such an attack, we consider 
its worst-case scenario. Fig.~\ref{fig:mfcc-exp} shows the evaluation 
flow. For each of the recorded command of the previous section, we 
extract the MFCCs for the full signal and use them to reconstruct 
the voice signal. Finally, we execute \name over the reconstructed 
voice segment and the corresponding \acc sample to test for a match.

We fixed the MFCC parameters as follows: 256 samples 
for the hop time, 512 samples for the window time, and 77
as the length of the output coefficient vector. We vary 
the number Mel filter bands between 15 and 30. At 15 Mel 
filter bands, the reconstructed voice command is similar 
to what is reported in existing attacks~\cite{vaidya2015cocaine}.
At 30 Mel filter bands, the reconstructed voice command
is very close to the original; it shares the same MFCCs 
and is easily identifiable when played back.

The question that we aim to address is whether 
reducing the sound signal to a set of features and 
reconstructing back the original signal
preserves all the acoustic features needed 
for \name to perform a successful matching with
the corresponding \acc signal. If not, then the 
reconstructed sound will not even match the voice
it originated from. Therefore, any mangled voice
will not match the user's speech as measured by 
\name, so that \name could successfully thwart the attack.

In all cases, while the original \mic signal matches
\acc signals near perfectly as indicated before,
the reconstructed sound failed to match the \acc
signal in 99\% of the evaluated cases. Of 3240 
comparisons (2 Mel filter band lengths per command, 90 
commands per user and 18 users), the reconstructed sound 
matched only a handful of \acc samples, and only in cases 
where we used 30 Mel filter bands. Indeed, those sound segments were very close to
the original sound segment that corresponds to the matched \acc samples.
To constitute an attack, the mangled voice segment is not supposed to originate 
from the sound the user is speaking, let alone preserving discernible 
acoustic features. This demonstrates that \name matches the time-domain signals in their 
entirety, thwarting such attacks on the voice assistant and recognition 
systems.

Last but not least, we tested \name with the set of mangled voice 
commands\footnote{http://www.hiddenvoicecommands.com/} used by 
Carlini \textit{et al.}~\cite{carlini2016hidden}. We asked 
four different individuals to repeat these commands while 
wearing \name. The \acc samples corresponding to each command 
do not match their mangled voice counterparts.

In scenario B, an attacker also fails to overcome \name's protection. 
We indicated earlier in Section~\ref{sec:phon_cross} and in this section that
\name successfully distinguishes 
the phonemes and commands of the same user.
We further confirm that \name can 
differentiate the same phoneme or command across different users. Moreover, 
even if the user is speaking and 
the adversary is replaying another sound clip of the same user, \name can
differentiate between the \mic and \acc samples and stop the attack.
Finally, \name might result in some false positives (albeit
very low). As explained earlier, these false
positive take place because the remaining segments after the 
per-segment stage of \name match, and thus
do not represent a viable attack vector. It is worth noting 
that \name could use a more stringent classifier that is tuned to force 
the false positive rate to be 0. This will come at the cost of usability
but could be preferable in high-security situations.

\subsection{Delay and Energy}

We measure the delay experienced at the voice assistant side and the energy 
consumption of the wearable component, using our prototype. 
As shown in Fig.~\ref{fig:arch}, \name incurs delay only during the matching phase: 
when \name uploads the \acc and \mic signals 
to the remote service and waits for a response. According to our test on the 
same list of 30 commands, we found that a successful match takes 300--830ms, 
with an average of 364ms, while an unsuccessful match takes 230--760ms, 
with an average of 319ms. The response time increases proportionally to the 
length of the commands, but matching a command containing more than 30 words
still takes less than 1 second. We expect the delay to decrease further if 
switching from our Matlab-based server to a full-fledged web server.

\begin{figure}
   \centering
   \includegraphics[width=0.75\columnwidth]{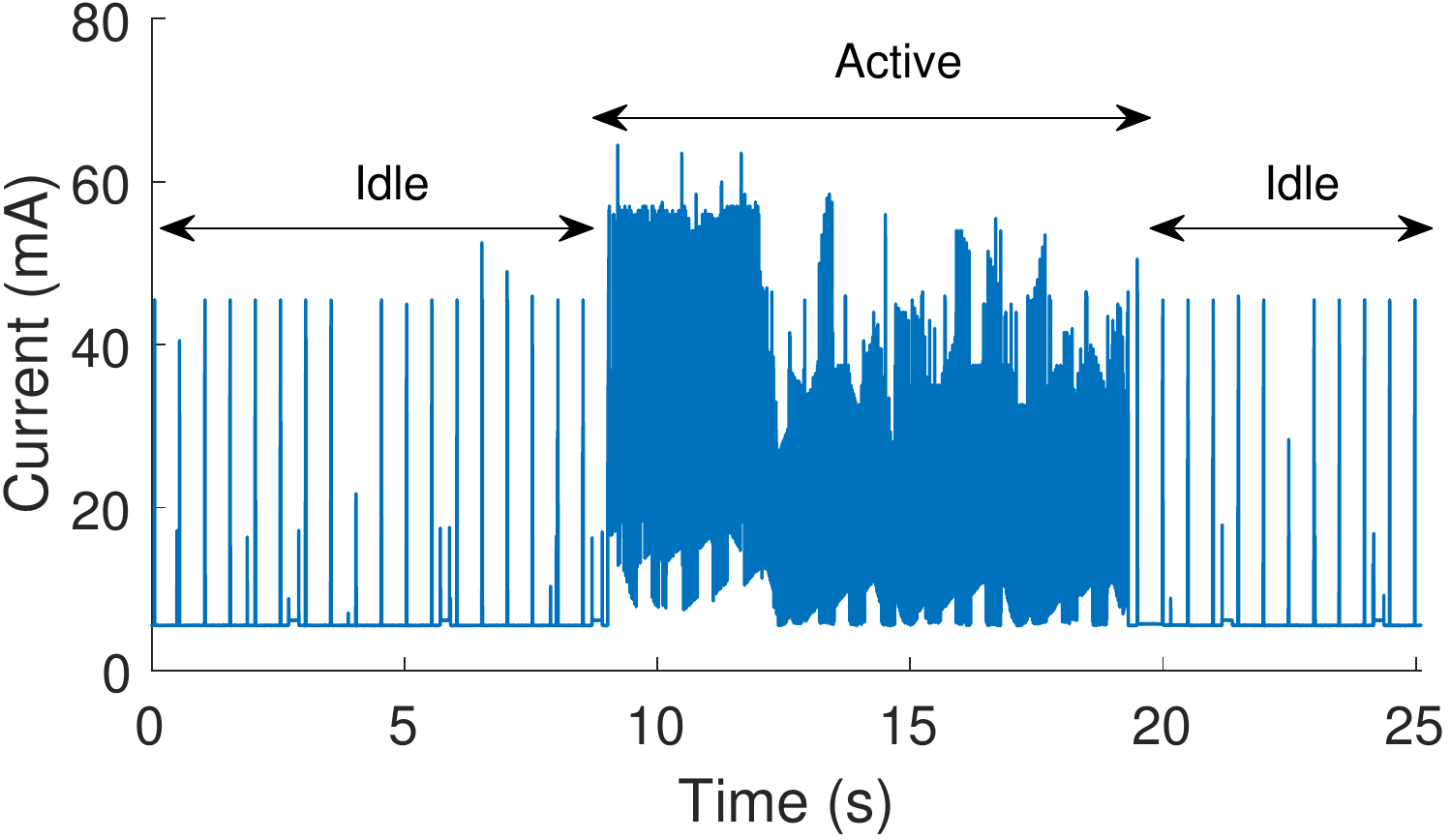}
   \caption{Current levels of the prototype in the idle and active states.}
   \label{fig:energy}
\end{figure}

When the wearable component transmits \acc signals, it switches between two 
states: idle state that keeps the connection alive and active state 
that actually transmits the data. We connected our prototype to 
the Monsoon power monitor and recorded the current levels of the prototype 
in these two states when powered by a fixed voltage (4V). Fig.~\ref{fig:energy} 
illustrates the changes of the current levels when our prototype switches from idle to 
active and then back to idle. We observed that under active state, our prototype 
consumes as much as 31mA, while under idle state, it only consumes an average 
of 6mA. Most of the energy is used to keep the Bluetooth connection and 
transmit data (in the active state) --- the energy consumed by the \acc sensor is almost negligible.

Assuming the user always keeps the wearable open at daytime and sends 100 voice 
commands per day (each voice command takes 10 seconds). Our prototype consumes 
6.3mA on average. This might even be an overestimation since 90\% of the 
users issue voice commands at most once per day according to our survey. 
A typical 500mAh Li-Ion battery used by wearables (comparable to a US quarter coin) 
can power our prototype for around a week. 80\% of the participants in our usability survey 
think they have no problem with recharging the wearable on a weekly basis. 
We conducted all the analyses on our prototype which directly utilizes off-the-shelf 
hardware chips without any optimization, assuming that \name is provided 
as a standalone wearable. If incorporated into an existing wearable device, 
\name will only introduce an additional energy overhead of less than 10mAh per day.


\section{Discussion}
\label{sec:discussion}

In our prototype implementation, we enforce the same policy for all voice commands: 
if the authentication passes, execute else drop the command. However, one can  
implement customized policy for different commands. For example, some commands, 
such as time/weather inquiry, are not privacy/security-sensitive, so \name can execute 
them directly without going through additional authentication process; other 
commands, such as controlling home appliances might be highly sensitive, and 
hence \name should promptly warn the user instead of simply dropping the command. 
This can be implemented by extending the Intent interception logic in our 
prototype implementation, making \name react differently according to 
different Intent actions, data, and types.

Besides its excellent security properties, \name has a distinct advantage 
over existing technologies --- it is wear-and-use without any user-specific, 
scenario-dependent training. Although we used machine learning to facilitate 
matching decision in our algorithm, we only trained once (on English phonemes 
of a test user) and then applied it in all other cases. Our evaluation demonstrates 
that \name is robust to changes in accents, speed of speech, mobility, 
or even languages. This significantly increases the usability of \name.


\section{conclusion}
\label{sec:conclusion}

In this paper, we have proposed \name, a system that provides 
continuous authentication for voice assistants. 
We demonstrated that even though the \acc information collected 
from the facial/neck/chest surfaces might be weak, it contains enough 
information to correlate it with the data received via microphone. 
\name provides extra physical assurance for voice assistant users and is an effective 
measure against various attack scenarios. It avoids the pitfalls of 
existing voice authentication mechanisms. 
Our evaluation with real users under practical settings shows high accuracy 
and very low false positive rate, highlighting the effectiveness of \name. 
In future, we would like to explore more configurations of \name that will promote 
wider real-world deployment and adoption.

\bibliographystyle{IEEEtran}
\bibliography{IEEEabrv,sigproc}

\newpage

\section*{Appendix A}
\label{appA}

Here, we list two tables which contain the English language phonetics as well as their examples
and the list of commands we utilized for evaluating \name.

\begin{table}[h]

\renewcommand{\arraystretch}{1}
\tiny
  \centering
   \caption[a]{The IPA chart of English phonetics.\footnotemark}
\begin{tabular}{C{0.8cm} C{2.1cm} | C{0.9cm} C{2cm}}
\toprule
\textbf{Vowel} & \textbf{Examples} & \textbf{Consonants} & \textbf{Examples}\\ 
\midrule 
\ipa{\:2}&C\textbf{U}P, L\textbf{U}CK &b&\textbf{B}AD, LA\textbf{B}\\
\ipa{\:A:}&\textbf{A}RM, F\textbf{A}THER &d&\textbf{D}I\textbf{D}, LA\textbf{D}Y\\
\ae&C\textbf{A}T, BL\textbf{A}CK &f&\textbf{F}IND, I\textbf{F}\\
e&M\textbf{E}T, B\textbf{E}D &g&\textbf{G}IVE, FLA\textbf{G}\\
\ipa{\:@}&AW\textbf{A}Y, CIN\textbf{E}M\textbf{A} &h&\textbf{H}OW, \textbf{H}ELLO\\
\ipa{\:3:$^r$}&T\textbf{UR}N, L\textbf{EAR}N &j&\textbf{Y}ES, \textbf{Y}ELLOW\\
\ipa{\:I}&H\textbf{I}T, S\textbf{I}TT\textbf{I}NG &k&\textbf{C}AT, BA\textbf{CK}\\
\ipa{\:i:}&S\textbf{EE}, H\textbf{EA}T &l&\textbf{L}EG, \textbf{L}ITTLE\\
\ipa{\:6}&H\textbf{O}T, R\textbf{O}CK &m&\textbf{M}AN, LE\textbf{M}ON\\
\ipa{\:O:}&C\textbf{A}LL, F\textbf{OU}R &n&\textbf{N}O, TE\textbf{N}\\
\ipa{\:U}&P\textbf{U}T, C\textbf{OU}LD &\ipa{\:N}&SI\textbf{NG}, FI\textbf{NG}ER\\
\ipa{\:u:}&BL\textbf{UE}, F\textbf{OO}D &p&\textbf{P}ET, MA\textbf{P}\\
\ipa{\:AI}&F\textbf{I}VE, \textbf{EYE} &r&\textbf{R}ED, T\textbf{R}Y\\
\ipa{\:AU}&N\textbf{OW}, \textbf{OU}T &s&\textbf{S}UN, MI\textbf{SS}\\
\ipa{\:eI}&S\textbf{A}Y, \textbf{EI}GHT &\ipa{\:S}&\textbf{SH}E, CRA\textbf{SH}\\
\ipa{\:oU}&G\textbf{O}, H\textbf{O}ME &t&\textbf{T}EA, GE\textbf{TT}ING\\
\ipa{\:OI}&B\textbf{OY}, J\textbf{OI}N &\ipa{\:tS}&\textbf{CH}ECK, \textbf{CH}UR\textbf{CH}\\
\ipa{\:e@$^r$}&WH\textbf{ERE}, \textbf{AIR} &\ipa{\:T}&\textbf{TH}INK, BO\textbf{TH}\\
\ipa{\:I@$^r$}&N\textbf{EAR}, H\textbf{ERE} &\ipa{\:D}&\textbf{TH}IS, MO\textbf{TH}ER\\
\ipa{\:U@$^r$}&P\textbf{URE}, T\textbf{OUR}IST&v&\textbf{V}OICE, FI\textbf{V}E\\
-&-&w&\textbf{W}ET, \textbf{W}INDO\textbf{W}\\
-&-&z&\textbf{Z}OO, LA\textbf{Z}Y\\
-&-&\ipa{\:Z}&PLEA\textbf{S}URE, VI\textbf{SI}ON\\
-&-&\ipa{\:dZ}&\textbf{J}UST, LAR\textbf{GE}\\

\bottomrule
\end{tabular} 

\label{table:phonetics}%
\end{table}%

\footnotetext{copied from: \url{http://www.antimoon.com/resources/phonchart2008.pdf}}
 
\begin{table}[h]

\renewcommand{\arraystretch}{1}
\tiny
  \centering
   \caption[a]{The List of commands.\footnotemark}
\begin{tabular}{L{4cm} | L{4cm}}
\toprule
\textbf{Command} & \textbf{Command} \\ 
\midrule

\textbf{1.} How old is Neil deGrasse Tyson? & \textbf{16.} Remind me to buy coffee at 7am from Starbucks \\
\textbf{2.} What does colloquial mean? & \textbf{17.} What is my schedule for tomorrow? \\
\textbf{3.} What time is it now in Tokyo? & \textbf{18.} Where's my Amazon package?  \\
\textbf{4.} Search for professional photography tips & \textbf{19.} Make a note: update my router firmware \\
\textbf{5.} Show me pictures of the Leaning Tower of Pisa & \textbf{20.} Find Florence Ion's phone number \\
\textbf{6.} Do I need an umbrella today? What's the weather like?  & \textbf{21.} Show me my bills due this week \\
\textbf{7.} What is the Google stock price? & \textbf{22.} Show me my last messages. \\
\textbf{8.} What's 135 divided by 7.5?  & \textbf{23.} Call Jon Smith on speakerphone \\
\textbf{9.} Search Tumblr for cat pictures & \textbf{24.} Text Susie great job on that feature yesterday \\
\textbf{10.} Open greenbot.com & \textbf{25.} Where is the nearest sushi restaurant? \\
\textbf{11.} Take a picture  & \textbf{26.} Show me restaurants near my hotel  \\
\textbf{12.} Open Spotify & \textbf{27.} Play some music  \\
\textbf{13.} Turn on Bluetooth & \textbf{28.} What's this song? \\
\textbf{14.} What's the tip for 123 dollars? & \textbf{29.} Did the Giants win today? \\
\textbf{15.} Set an alarm for 6:30 am & \textbf{30.} How do you say good night in Japanese? \\

\bottomrule
\end{tabular} 

\label{table:commands}%
\end{table}%

\footnotetext{inspired from: \url{http://www.greenbot.com/article/2359684/android/a-list-of-all-the-ok-google-voice-commands.html}}

\end{document}